\documentclass[%
amsmath,amssymb,
aps,
twocolumn,
]{revtex4-2}

\usepackage{graphicx}
\usepackage{dcolumn}
\usepackage{bm}

\usepackage{physics}
\usepackage{amsmath,amssymb}
\usepackage{tikz}
\usepackage{geometry}
\usepackage{hyperref}

\hypersetup{colorlinks=true}

\begin{document}
	
	\preprint{APS/123-QED}
	
	\title{Searching for the first radial excitation of the $ \Delta(1232) $ in lattice QCD}%
	\author{Liam Hockley}
	\email{liam.hockley@adelaide.edu.au}
	\author{Waseem Kamleh}%
	\author{Derek Leinweber}
	\author{Anthony Thomas}
	\affiliation{%
		ARC Special Research Centre for the Subatomic Structure of Matter (CSSM), Department of Physics, The University of Adelaide, SA, 5005,
		Australia.
	}%

	\date{\today}
	
	\begin{abstract}
		We present a lattice QCD analysis of the $ \Delta $-baryon spectrum, with the goal of finding the position of the $ 2s $ radial excitation of the $ \Delta(1232) $ ground state. Using smeared three-quark operators in a correlation matrix analysis, we report masses for the ground, first and second excited states of the $ J^P = 3/2^+ $ spectrum across a broad range of $ m_\pi^2 $. We identify the lowest lying state as being a $ 1s $ state, consistent with the well known $ \Delta(1232) $. The first excitation is identified as a $ 2s $ state, but is found to have a mass of approximately 2.15~GeV on our $ \sim3 $ fm lattice, which does not appear to be associated with the $ \Delta(1600) $ resonance in a significant manner. We also report on the spin-$ 1/2 $ and odd-parity states accessible via our methods. The large excitation energies of the radial excitations provide a potential resolution to the long-standing missing baryon resonances problem.

	\end{abstract}
	
	
	\maketitle

	\section{Introduction}\label{sec:Intro}
	
	Short-lived hadronic states, otherwise known as resonances, present various challenges in our understanding of the strong force. While the quark model performs well for describing long-lived ground state mesons and baryons \cite{Gell-Mann:1961omu,Gell-Mann:1962yej}, the complex dynamics of resonances and the possibility of several chains of decays makes modelling and analysis of these and their excited states non-trivial. The foremost example is the $ N^*(1440) $ Roper resonance \cite{Roper:1964zza, Krehl:1999km, Burkert:2017djo}, with similar issues associated with the $ \Delta(1600) $ resonance \cite{Liu:2022ndb, Ramalho:2010cw}. A recent description, arising through Dyson-Schwinger methods, is that the first excitation in both of these channels is a radial excitation of the ground state \cite{Burkert:2017djo,Liu:2022ndb}.
	
	The only currently tenable method for resolving the properties of QCD at hadronic energy scales is lattice QCD. Lattice QCD is well established as a non-perturbative framework within which many physical quantities can be extracted such as particle masses, decay constants, form factors, etc. The systematics to do with the discretisation of spacetime and using heavier-than-physical quark masses are now readily controlled and understood, leading to sub-percent level errors \cite{FLAG:2021npn}.
	
	Light baryon resonances are of particular interest in the context of lattice QCD studies for several reasons. Firstly, as energy increases beyond relevant thresholds, state of the art calculations need to consider contributions from a number of open many-body channels, as in Refs. \cite{Bulava:2022vpq, Alexandrou:2023elk, Bulava:2023uma}. This necessitates a robust basis of interpolators that couple to these scattering states, and so provide a spectrum of energies which can be connected to infinite-volume resonances by methods such as that of L\"{u}scher \cite{Luscher:1985dn,Luscher:1986pf,Luscher:1990ux}. 
	
	One such approach, known as Hamiltonian Effective Field Theory (HEFT), has been implemented in the case of the Roper resonance, leading to strong evidence that it is not a simple radial excitation of the ground state nucleon \cite{Mahbub:2010rm,Kiratidis:2015vpa, Kiratidis:2016hda,Liu:2016uzk,Wu:2017qve}. This approach draws on lattice QCD finite-volume energies across a broad range of $ m_\pi^2 $, and so, in anticipation of applying this formalism to extract the nature of the $ \Delta(1600) $ resonance, the current work aims to provide a similar foundation of masses in the $ \Delta $ regime.
	
	In studying the Roper, it was found that a basis of smeared three-quark interpolating fields provides a method of classifying the associated eigenstates in terms of radial excitations \cite{Mahbub:2013ala}, based on the number of nodes in the wave function \cite{Roberts:2013oea, Roberts:2013ipa}. We apply the same approach here in the $ \Delta $ spectrum, and seek to identify the approximate position of the first radial excitation of the $ \Delta(1232) $. This will provide some initial insight into the nature of the $ \Delta(1600) $, and whether it is indeed the radial excitation of the ground state $ \Delta $.
	
	In constructing a basis of three-quark interpolating fields, one finds that various combinations of total angular momentum and parity can be projected out, and so it is of interest to also explore beyond the $ \Delta(3/2^+) $ and consider the $ 3/2^-,\, 1/2^+ $ and $ 1/2^- $ states in the $ \Delta $ spectrum. Resonances of interest, as well as their PDG classification in terms of how well studied they are, are given in Table~\ref{tab:DeltaPDG}.
	
	\begin{table} [b]
		\begin{center}
			\renewcommand{\arraystretch}{1.5}
			\begin{tabular}{ccr}
				\hline
				\hline
				$ m $ (MeV) & $ J^P $ & Classification  \\
				\hline
				$ 1232 $ & $ 3/2^+ $ & **** \\
				$ 1600 $ & $ 3/2^+ $ & **** \\
				$ 1620 $ & $ 1/2^- $ & **** \\
				$ 1700 $ & $ 3/2^- $ & **** \\
				$ 1750 $ & $ 1/2^+ $ & *    \\
				$ 1900 $ & $ 1/2^- $ & ***  \\
				$ 1910 $ & $ 1/2^+ $ & **** \\
				$ 1920 $ & $ 3/2^+ $ & ***  \\
				$ 1940 $ & $ 3/2^- $ & **   \\
				$ 2150 $ & $ 1/2^- $ & *   \\
				\hline
			\end{tabular}
			\caption{Table of resonances in the $ \Delta $ spectrum ($ J^P = 1/2^\pm $ and $ 3/2^\pm $ only) based on the PDG tables \cite{ParticleDataGroup:2022pth}. Classification scheme: 
				****   	Existence is certain, and properties are at least fairly explored.
				***   	Existence ranges from very likely to certain, but further confirmation is desirable and/or quantum numbers, branching fractions, etc. are not well determined.
				**   	Evidence of existence is only fair.
				*   	Evidence of existence is poor.}
			\label{tab:DeltaPDG}
		\end{center}
	\end{table}

	The paper is organised as follows. Section~\ref{sec:Lattice} describes the methodology used to compute the $ \Delta $ spectrum. We outline the variational method used for extracting excited states in the spectrum as well as a spin projection technique. Finally, we present techniques used for identifying radial excitations in the spectrum based on nodes generated in the superposition of interpolating fields. Our results for the $ J^P = 3/2^+ $ spectrum are discussed in detail in Section~\ref{sec:Results1}. While our focus is on the $ 3/2^+ $ spectrum, Section~\ref{sec:Results2} contains results for the $ 3/2^- $ and $ 1/2^\pm $ states that we are also able to extract using our methods. These act to disclose the role of wave function nodes in the excitations, with additional comparison available from the work of other recent studies.
	
	\section{Lattice QCD Calculations}\label{sec:Lattice}
	In this work we use well established tools for extracting baryon spectra within the framework of lattice QCD. This involves constructing interpolating fields which couple to single-particle, three-quark dominated states on the lattice with quantum numbers of the $ \Delta $-baryon, and constructing 2-point correlation functions from these. These correlation functions can then be spin- and parity-projected to isolate different states sharing the same quark flavours, whose masses can then be extracted using the variational method \cite{Michael:1985ne,Luscher:1990ck,Mahbub:2013ala}. The details of each of these steps will be explored in the subsections below.
	
	\subsection{Baryon Spectroscopy}\label{subsec:BarSpec}	
	The masses of states in the energy spectrum of a particular baryon species can be simply derived from 2-point correlation functions.	At the hadronic level, a general 2-point correlation function is defined as
	\begin{equation}
		\mathcal{G}^{ij}_{\mu\nu}(t,\boldsymbol{p}) = \sum_{\boldsymbol{x}} e^{-i\boldsymbol{p}\cdot \boldsymbol{x}} \mel{\Omega}{T\{ \chi^i_\mu(x) \bar{\chi}^j_\nu(0) \}}{\Omega}\,.
	\end{equation}
	The indices $ i, j $ refer to choices of interpolating field $ \chi $, with Greek indices $ \mu,\nu $ labelling the Lorentz degrees of freedom. $ T $ indicates time ordering of the operators. Putting all this together, the operator $ \bar{\chi}^j_\nu(0) $ acts on the vacuum to create states at the space-time point $ (0, \boldsymbol{0}) $. These states then propagate through Euclidean time $ t $ before being annihilated at a new space-time point $ x = (t,\boldsymbol{x}) $ by the operator $ \chi^i_\mu(x) $. The Fourier transform projects the baryon momentum $ \boldsymbol{p} $.
	
	For the case of a particular baryon species, we have the complete set of states
	\begin{equation}
		\sum_{B,\boldsymbol{p}\,',s} \ket{B,\boldsymbol{p}\,',s}\bra{B,\boldsymbol{p}\,',s} = I\,,
	\end{equation}
	where $ B $ labels the baryon species with momenta $ \boldsymbol{p}' $ and spins $ s $. This allows us to obtain
	\begin{align}
		\mathcal{G}^{ij}_{\mu\nu}(t,\boldsymbol{p}) 
		= \sum_{B,\boldsymbol{p}\,',s} \sum_{\boldsymbol{x}}& 
		e^{-i\boldsymbol{p}\cdot \boldsymbol{x}} 
		\mel{\Omega}{\chi^i_\mu(x)}{B,\boldsymbol{p}\,',s} \nonumber \\
		& \times \mel{B,\boldsymbol{p}\,',s}{\bar{\chi}^j_\nu(0)}{\Omega}\,.
	\end{align}

	In order to have meaningful overlap with this complete set of baryon states, the form of the interpolating fields should be chosen so that they share the same quantum numbers and symmetry properties as the states we wish to extract. 
	
	We then use the translational operator to rewrite $ \chi^i_\mu(x) $ as
	\begin{equation}
		\chi^i_\mu(x) = e^{Ht}\, e^{-i\boldsymbol{P}\cdot\boldsymbol{x}}\, \chi^i_\mu(0)\, e^{i\boldsymbol{P}\cdot\boldsymbol{x}}\, e^{-Ht}\,.
	\end{equation}
	This allows us to simplify the expression for the correlation function to
	\begin{align}
		\mathcal{G}^{ij}_{\mu\nu}(t,\boldsymbol{p}) 
		= \sum_{B,s} &
		e^{-E_Bt} 
		\mel{\Omega}{\chi^i_\mu(0)} {B,\boldsymbol{p},s} \nonumber \\ 
		&\times \mel{B,\boldsymbol{p},s}{\bar{\chi}^j_\nu(0)}{\Omega}\,. \label{eq:correlator}
	\end{align}
	
	At this point it becomes important to distinguish the states allowed for the system in question. Given our choice of three-quark operator for the interpolating field coupling to the $ \Delta^{++} $
	\begin{equation}
		\chi_\mu(x) = \qty(\bar{u}(x) C \gamma_\mu u(x)) u(x)\,,
	\end{equation}
	we need to consider both spin-3/2 and spin-1/2 contributions. 
	
	\begin{widetext}
		
		Thus, for our case of interest we have the following available matrix elements, with the associated decomposition into Dirac spinors $ u(\boldsymbol{p},s) $ for the spin-1/2 components, and Rarita-Schwinger spinors $ u_\mu(\boldsymbol{p},s) $ for the spin-3/2 components:
		
		\begin{align}
			&\mel*{\Omega}{\chi^i_\mu(0)}{\Delta^{3/2^+}(\boldsymbol{p},s)} = \lambda_{3/2^+} \sqrt{\frac{M_{3/2^+}}{E_{3/2^+}}}\, u_\mu (\boldsymbol{p},s)\,, \label{eq:matprod1} \\
			&\mel*{\Omega}{\chi^i_\mu(0)}{\Delta^{3/2^-}(\boldsymbol{p},s)} = \lambda_{3/2^-} \sqrt{\frac{M_{3/2^-}}{E_{3/2^-}}}\, \gamma_5 \, u_\mu (\boldsymbol{p},s)\,, \label{eq:matprod2} \\
			&\mel*{\Omega}{\chi^i_\mu(0)}{\Delta^{1/2^+}(\boldsymbol{p},s)} = \qty(\alpha_{1/2^+}p_\mu + \beta_{1/2^+}\gamma_\mu)
			\sqrt{\frac{M_{1/2^+}}{E_{1/2^+}}} \, \gamma_5 \, u (\boldsymbol{p},s)\,, \label{eq:matprod3} \\
			&\mel*{\Omega}{\chi^i_\mu(0)}{\Delta^{1/2^-}(\boldsymbol{p},s)}= \qty(\alpha_{1/2^-}p_\mu + \beta_{1/2^-}\gamma_\mu) 
			\sqrt{\frac{M_{1/2^-}}{E_{1/2^-}}}\, u(\boldsymbol{p},s)\,. \label{eq:matprod4} 
		\end{align}
		Here, the $ \lambda $, $ \alpha $ and $ \beta $ factors are various couplings and $ M $ is the mass of the state with relativistic energy $ E $. We note that as both the Rarita-Schwinger spinor and the interpolating field transform under parity as a pseudovector, Eq.~\eqref{eq:matprod3} includes a $ \gamma_5 $ next to the Dirac spinor. Including the corresponding expressions for the $ \bar{\chi} $ matrix elements from Eq.~(\ref{eq:correlator}), one can use the Dirac spinor spin-sum relation
		\begin{equation}
			\sum_{s = -1/2}^{1/2} u(\boldsymbol{p},s) \bar{u}(\boldsymbol{p},s) = \frac{\gamma \cdot p + M_B}{2M_B}\,, \label{eq:diracspinsum}
		\end{equation}
		and the analogous Rarita-Schwinger spin-sum relation 
		\begin{align}
			&\sum_{s = -3/2}^{3/2} u_\mu(\boldsymbol{p},s) \bar{u}_\nu(\boldsymbol{p},s) = -\frac{\gamma \cdot p + M_B}{2M_B} \qty{ g_{\mu\nu} - \frac{1}{3}\gamma_\mu \gamma_\nu - \frac{2p_\mu p_\nu}{3M_B^2} + \frac{p_\mu \gamma_\nu - p_\nu \gamma_\mu}{3M_B} }\,, \label{rsspinsum}
		\end{align}
		to compute the sum over $ s $ and write the contributions from each of the required products of matrix elements. At $ \boldsymbol{p} = 0 $, so that $ E_{B}~=~M_{B} $, and with $ \mu = \nu = n = 1,2,3 $, we have
		\begin{align}
			&\mel*{\Omega}{\chi^i_n(0)}{\Delta^{3/2^+}(\boldsymbol{p},s)} \mel*{\Delta^{3/2^+}(\boldsymbol{p},s)}{\bar{\chi}^j_n(0)}{\Omega} = \lambda^i_{3/2^+}\, \overline{\lambda}^{j}_{3/2^+}\, \frac{2}{3} \qty( \frac{\gamma_0 + \mathbb{I}}{2} ) \\
			&\mel*{\Omega}{\chi^i_n(0)}{\Delta^{3/2^-}(\boldsymbol{p},s)} \mel*{\Delta^{3/2^-}(\boldsymbol{p},s)}{\bar{\chi}^j_n(0)}{\Omega} = \lambda^i_{3/2^-}\, \overline{\lambda}^{j}_{3/2^-}\, \frac{2}{3} \qty( \frac{\gamma_0 - \mathbb{I}}{2} ) \\
			&\mel*{\Omega}{\chi^i_n(0)}{\Delta^{1/2^+}(\boldsymbol{p},s)} \mel*{\Delta^{1/2^+}(\boldsymbol{p},s)}{\bar{\chi}^j_n(0)}{\Omega} = \lambda^i_{1/2^+}\, \overline{\lambda}^{j}_{1/2^+}\, \qty( \frac{\gamma_0 + \mathbb{I}}{2} ) \\
			&\mel*{\Omega}{\chi^i_n(0)}{\Delta^{1/2^-}(\boldsymbol{p},s)} \mel*{\Delta^{1/2^-}(\boldsymbol{p},s)}{\bar{\chi}^j_n(0)}{\Omega} = \lambda^i_{1/2^-}\, \overline{\lambda}^{j}_{1/2^-}\, \qty( \frac{\gamma_0 - \mathbb{I}}{2} )\,.
		\end{align}
		where we've relabelled $ \beta_{1/2^\pm} \to \lambda_{1/2^\pm} $ for simplicity. The notation $ \overline{\lambda} $ is used to denote the overlap of the interpolator at the source.
		
	\end{widetext}
		
	Hence we can see that the even-parity states exist in the $ (1,1) $ and $ (2,2) $ Dirac components of the correlation function, while the odd-parity states live in the $ (3,3) $ and $ (4,4) $ Dirac components. These are then readily accessed by applying the appropriate projection operators	
	\begin{equation}
		\Gamma^\pm = \frac{1}{2} \qty(\gamma_0 \pm \mathbb{I})\,. \label{eq:ParProj}
	\end{equation}
	and taking the trace over Dirac indices.
	
	In particular, we note that the parity information of our correlation functions is governed by the upper and lower components of the spinors describing the baryon states. This will be of importance in Section~\ref{sec:Results2} in discussing the identification of radial excitations for odd-parity states.
	
	One can then define the parity projected correlator as a sum over the spatial Lorentz indices $ \mu = \nu = n $ \cite{Leinweber:1992hy}
	\begin{equation}
		G^\pm_{ij}(t,\boldsymbol{0}) \equiv \tr_{sp} \qty[\Gamma^\pm \sum_n \mathcal{G}^{ij}_{nn}(t,\boldsymbol{0})]\,.  \label{eq:trace}
	\end{equation}
	
 	The result is a series of decaying exponentials governed by the mass of the baryon states
	\begin{equation}
		G^\pm_{ij}(t,\boldsymbol{0}) = \sum_{B^\pm} \lambda_{iB^\pm}\, \overline{\lambda}_{jB^\pm}\, e^{-M_{B^\pm}t}\,, \label{trace_corr}
	\end{equation}
	where $ \lambda_{i B^\pm} $ and $ \overline{\lambda}_{j B^\pm} $ are coupling strengths between the interpolating fields $ \chi^i_\mu $ and $ \overline{\chi}^j_\nu $ and the parity projected baryon states $ B^{\pm} $, up to some overall constant factors. We note that these coupling strengths can be taken to be real by considering both the original gauge-field links and their complex conjugates, weighted equally in the ensemble average \cite{Leinweber:1992hy, Mahbub:2013ala}.
	
	The parity projected correlation function still contains various spin states (such as spin-1/2 and spin-3/2 in our case) as well as a tower of excited states. In order to extract, say, the ground state mass of a particular baryon state, one takes the long-time limit in which all the excited states have decayed off. Explicitly,
	\begin{equation}
		G^\pm_{ij}(t,\boldsymbol{0}) \overset{t\to\infty}{=} \lambda^\pm_{i0}\, \overline{\lambda}^\pm_{j0}\, e^{-M_{0^\pm}t}\,, \label{groundcorr}
	\end{equation}
	where the $ \lambda^\pm_{i0} $ and $\overline{\lambda}^\pm_{j0} $ are couplings of baryon interpolators at the source and sink to the lowest lying state. In order to isolate spin states one can employ further projection techniques as discussed in the following section. A discussion of how one can access excited energy states is presented in Section~\ref{subsec:VarMeth}.
	
	\subsection{Interpolating Operators and Spin Projection} \label{subsec:IntOps}
	To simulate $ \Delta $ baryons on the lattice, we start from the well established form for the $ \Delta^{++} $ \cite{Ioffe:1981kw, Chung:1981cc}
	\begin{equation}
		\Delta^{++}_\mu (x) = \epsilon^{abc} \qty(u^{Ta}(x) C\gamma_\mu u^{b}(x)) u^{c}(x) \label{eq:deltainterp}\,.
	\end{equation}	
	The superscript $ T $ denotes the transpose, $ a,b,c $ are colour indices and $ u(x) $ is the Dirac spinor field for the $ u $-quark. Note that this interpolating field couples to both spin-1/2 and spin-3/2 states (by coupling of the spins of three spin-1/2 quarks). Hence we require additional spin projection to distinguish the relevant contributions to the overall correlation function.
	
	The general form of a spin-$ s $ projected correlator $ G^s $ in our formalism is
	\begin{equation}
		\mathcal{G}^s_{\mu\nu} = \sum_{\sigma,\lambda = 1}^{4} \mathcal{G}_{\mu\sigma}\, g^{\sigma\lambda}\, P^s_{\lambda\nu}\,,
	\end{equation}
	where $ P^s_{\mu\nu} $ are spin-$ s $ projectors for $ s = 1/2 $ or $ 3/2 $. The form of these is given in Appendix~\ref{app:SpinProj}. Once a correlator is spin-projected, it can then be subsituted directly into Eq.~(\ref{eq:trace}) with the calculation proceeding as shown.
			
	\subsection{Source and Sink Smearing}\label{subsec:SourSinkSmear} 
\begin{figure*}[]
		\includegraphics[width=0.3\linewidth]{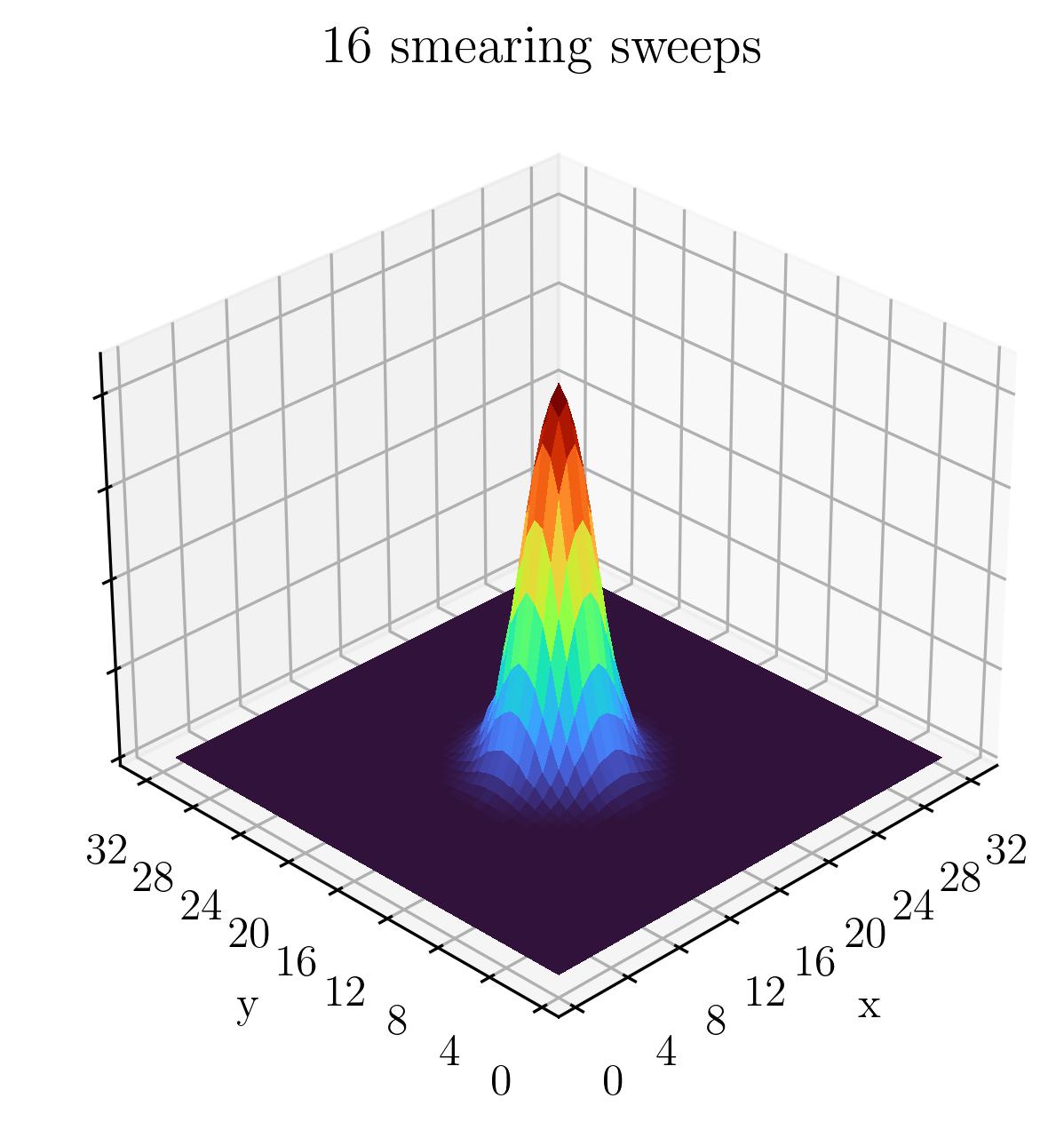}\hfil
		\includegraphics[width=0.3\linewidth]{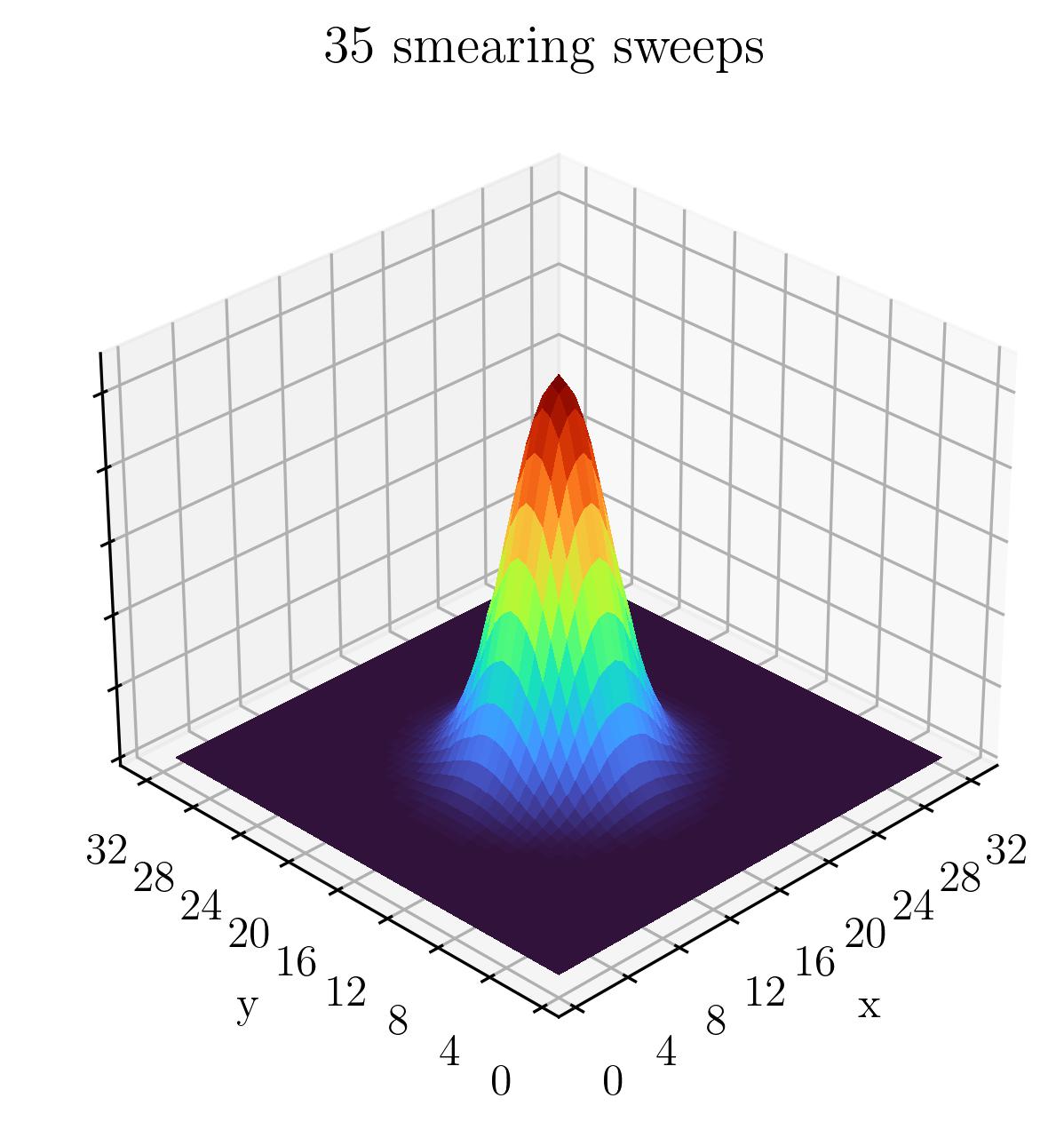}
		
	    \includegraphics[width=0.3\linewidth]{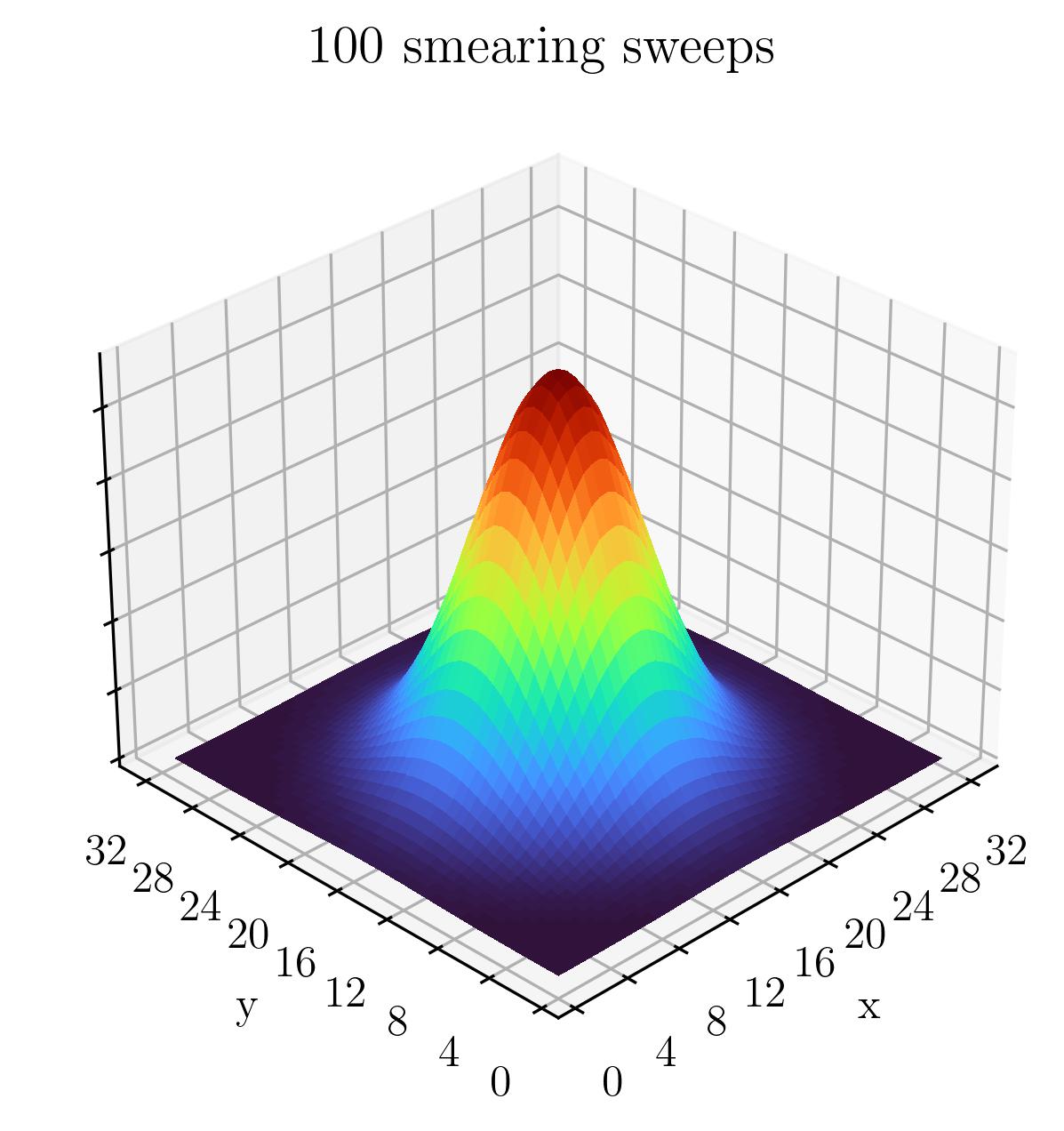}\hfil
		\includegraphics[width=0.3\linewidth]{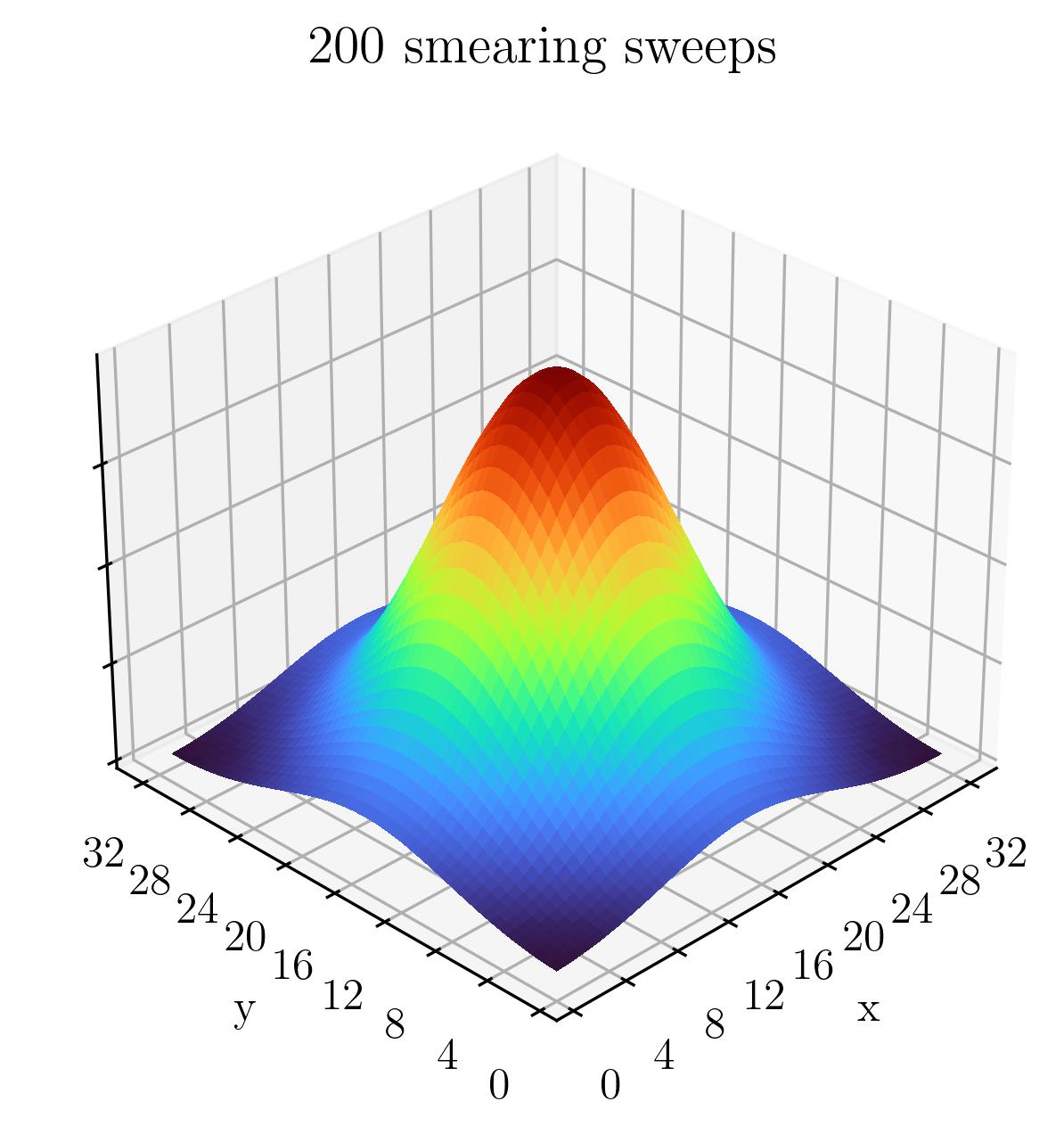}
		\caption{Plots of point sources after applying the smearing operator 16, 35, 100 and 200  times. The height of the peak is kept fixed in the visualisation to illustrate the broadening of the quark distribution.}
		\label{fig:SmearGauss}
	\end{figure*}

	In order to improve the overlap of the interpolating fields with the states of interest, we apply Gaussian smearing to the spatial components of the interpolating fields. The general procedure is to take some fermion field $ \psi_i(t, \boldsymbol{x}) $ and iteratively apply a smearing function $ F(\boldsymbol{x},\boldsymbol{x}') $. Explicitly, this takes the form
	\begin{equation}
		\psi_i(t,\boldsymbol{x}) = \sum_{\boldsymbol{x}'}\, F(\boldsymbol{x},\boldsymbol{x}')\, \psi_{i-1}(t,\boldsymbol{x}')
	\end{equation}
	where the smearing function is
	\begin{align}
		F(\boldsymbol{x},\boldsymbol{x}') &= (1-\alpha)\,\delta_{x,x'} \nonumber
		\\
		&+ \frac{\alpha}{6} \sum_{\mu = 1}^{3} \qty[U_\mu(x)\,\delta_{x',x+\hat{\mu}} + U^\dagger_\mu(x - \hat{\mu})\,\delta_{x',x-\hat{\mu}}]\,.
	\end{align}
	We take the smearing parameter to be $ \alpha=0.7 $ in our calculations. The use of repeated applications of the smearing function controls the width of our source by gradually smearing out an initial point source. This is visualised in Fig.~\ref{fig:SmearGauss} where, for 16, 35, 100 and 200 smearing sweeps, the amplitude of the distribution for $ U_\mu(x) = \mathbb{I} $ is plotted on the x-y plane with the third spatial dimension fixed at the centre of the source. These choices of smearing sweeps are motivated by considering the condition number of the correlation matrix, as in Ref.~\cite{Mahbub:2013ala}.
	
	\subsection{Variational Method}\label{subsec:VarMeth} 
	Recall that in Section~\ref{subsec:BarSpec} we showed how one can readily obtain the mass of baryon ground states by simply taking the leading order contribution to the correlation function as in Eq.~\eqref{groundcorr}. In order to isolate states of higher energies, a more nuanced approach is required, since these states are at sub-leading order in the exponential series. We make use of the well-known variational analysis method \cite{Michael:1985ne, Mahbub:2013ala} in order to extract the ground state mass as well as the excited state masses.
	
	To extract $ N $ states, we require $ N $ interpolating fields (one for each state). We generate our interpolating fields by using the same base field for the $ \Delta^+ $ baryon, obtained from Eq.~(\ref{eq:deltainterp}) and applying different levels of smearing consisting of 16, 35, 100 or 200 sweeps of smearing. These smearing levels are chosen both for their coupling to the states of interest and to produce a well defined matrix of correlation functions after applying these smearing levels to both the source and sink \cite{Mahbub:2010rm, Mahbub:2013ala}. This correlation matrix is written as
	
	\begin{equation}
		G^\pm_{ij}(t) = \sum_{\alpha=0}^{N-1}\, \lambda^\alpha_i \, \overline{\lambda}^\alpha_j \, e^{-m_\alpha t}\,. \label{corr3}
	\end{equation}
	Here the $ \lambda^\alpha_i $ and $ \overline{\lambda}^\alpha_j $ are essentially the same as seen before in Section~\ref{subsec:BarSpec}, though we now use the $ \alpha $ index to distinguish between energy states. In other words, they are couplings of the smeared interpolators $ \chi^i_\mu $ and $ \overline{\chi}^j_\nu $ at the sink and source, respectively, to the various energy eigenstates $ \alpha = 0, \dots , N-1 $. $ m_\alpha $ is the mass of the state $ \alpha $.
	
	From here, we now aim to construct linear combinations of our smeared interpolating fields to cleanly isolate the $ N $ states in the baryon spectrum. Labelling these baryon states $ \ket{B_\alpha} $, we thus wish to construct the superpositions
	\begin{equation}
		\overline{\phi}^\alpha_\mu 	= \sum_{i = 1}^{N} u^\alpha_i\, \overline{\chi}^i_\mu\,, \qquad \phi^\alpha_\nu = \sum_{i = 1}^{N} v^\alpha_i\, \chi^i_\nu\,, \label{phibar}
	\end{equation}
	such that
	\begin{align}
		\mel*{B_\beta, \boldsymbol{p},s}{\overline{\phi}^\alpha_\mu}{\Omega} &= \delta_{\alpha\beta}\, \overline{z}^\alpha\, \overline{u}_\mu(\alpha,\boldsymbol{p},s)\,, \\
		\mel*{\Omega}{\phi^\alpha_\nu}{B_\beta, \boldsymbol{p},s} &= \delta_{\alpha\beta}\, z^\alpha\, u_\nu(\alpha,\boldsymbol{p},s)\,, \label{phi}
	\end{align}
	where $ u_\mu(\alpha, \boldsymbol{p},s) $ is a Rarita-Schwinger spin vector. Here, the $ z^\alpha $ and $ \overline{z}^\alpha $ are the couplings of the superpositions $ \phi^\alpha_\mu $ and $ \overline{\phi}^\alpha_\nu $ to the state $ \ket{B_\alpha} $. The $ u^\alpha_i $ and $ v^\alpha_i $ are simply the weights for the basis of smeared interpolating fields.
	
	At this point, we construct an eigenvalue problem to solve for both $ \boldsymbol{u}^\alpha $ and $ \boldsymbol{v}^\alpha $. Noting that since $ G_{ij}(t) $ is real and symmetric, $ G_{ij}(t) = G_{ji}(t) $, in the ensemble average we introduce an improved unbiased estimator of the correlation matrix $ \qty[G_{ij}(t) + G_{ji}(t)]/2 $. This provides us with a correlation matrix which is symmetric, so we can simultaneously compute $ \boldsymbol{u}^\alpha $ and $ \boldsymbol{v}^\alpha $ as discussed below.
	
	Multiplying Eq.~\eqref{corr3} on the right by $ u^\alpha_j $ we obtain
	\begin{equation}
		G^\pm_{ij}(t)\, u^\alpha_j = \lambda^\alpha_i \, \overline{z}^\alpha\, e^{-m_\alpha t}\,.
	\end{equation}

	Then, since the exponential is the only time dependent part of the correlation function, we can form a  recurrence relation at some time after source insertion by introducing the variational parameters $ t_0 $ and $ \Delta t $:
	\begin{equation}
		G_{ij}(t_0 + \Delta t)\, u^\alpha_j = e^{-m_\alpha \Delta t}\, G_{ij}(t_0)\, u^\alpha_j\,.
	\end{equation}
	
	Then, multiplying on the left by the inverse $ [G_{ij}(t_0)]^{-1} $ and suppressing the indices $ i $ and $ j $ gives 
	\begin{equation}
		[G(t_0)^{-1} G(t_0 + \Delta t)]\, \boldsymbol{u}^\alpha = e^{-m_\alpha \Delta t}\,\boldsymbol{u}^\alpha\,, \label{genu}
	\end{equation}
	which we recognise as an eigenvalue equation for the vector in interpolator space $ \boldsymbol{u}^\alpha $.
	
	Similarly, premultiplying Eq.~\eqref{corr3} by $ v^\alpha_i $ we get
	\begin{equation}
		v^\alpha_i G_{ij}\,(t_0 + \Delta t) = e^{-m_\alpha \Delta t}\, v^\alpha_i \, G_{ij}(t_0)
	\end{equation}
	from which we arrive at our second eigenvalue equation (this time for $ \boldsymbol{v}^\alpha $):
	\begin{equation}
		\boldsymbol{v}^\alpha\, G(t_0 + \Delta t)\,[G(t_0)]^{-1} = e^{-m_\alpha \Delta t}\, \boldsymbol{v}^\alpha\,. \label{genv}
	\end{equation}
	
	Both Eq.~\eqref{genu} and \eqref{genv} need to be solved simultaneously for each given pair of variational parameters $ t_0 $ and $ \Delta t $, and we do so using a generalised eigenvalue problem solver. Solving for these eigenvectors automatically gives us the weights for the superpositions of interpolating fields, by construction. 
	
	Finally, the eigenstate and parity projected correlation function is then taken to be
	\begin{equation}
		G^\alpha_\pm \equiv v^\alpha_i \, G^\pm_{ij}(t)\, u^\alpha_j\,.
	\end{equation}
	The eigenvectors are essentially used to isolate particular states in the baryon spectrum, exactly as we set out to do. We then construct the effective mass function
	\begin{equation}
		M^\alpha_{\text{eff}}(t) = \frac{1}{\delta t} \ln(\frac{G^\alpha_\pm(t,\boldsymbol{0})}{G^\alpha_\pm(t+\delta t,\boldsymbol{0})})\,. \label{effmass}
	\end{equation}
	$ \delta t $ is typically taken to be small and is set independently of the variational parameters. We take $ \delta t = 2 $ in our calculations.
	
	It is also worth noting that the eigenvectors $ \boldsymbol{u}^\alpha $ and $ \boldsymbol{v}^\alpha $ are equal since $ G_{ij}(t) $ is a real symmetric matrix. From here on, we will refer to the $ \boldsymbol{u}^\alpha $ vector, for simplicity.
	
	The effective mass defined in Eq.~\eqref{effmass} can be computed for various discrete values of $ t $ and then plotted as a function of Euclidean time. As usual, one looks for time intervals over which the effective mass plot plateaus, indicating that all contamination from unwanted excited states has decayed away in the exponential series. We then perform a covariance matrix analysis of the $ \chi^2 $ per degree of freedom \cite{Mahbub:2013bba} to determine the most suitable time intervals to fit when obtaining our final masses. The details of this procedure for calculating our fit uncertainties is given in Appendix~\ref{app:jack}.
	
	With a basis of 4 smeared interpolating fields, we have access to 4 energy eigenstates in principle. However, while the third excited state is available to us, it is susceptible to excited state contamination. In some cases the signal-to-noise ratio degrades too rapidly for us to reliably report results, and these are omitted. Hence we only report the masses of the ground, first and second excited states.
	
	As a final note, the principles highlighted within this section are fully realised only when one includes a complete set of interpolating fields effective at isolating all the states within the spectrum. In principle this needs to include multi-particle scattering states as highlighted in, for example, Ref.~\cite{Hansen:2019nir}. However, our aim is more modest; to determine the approximate position of the $ 2s $ excitation of the $ \Delta(1232) $ in the spectrum. Our formalism is suited to exciting these single-particle states and in particular identifying their radial node structure. As such, Euclidean time evolution is important in suppressing contamination from nearby states. In fitting our effective mass to a plateau, we enhance single-state dominance by monitoring the $ \chi^2 $ per degree of freedom ($ \chi^2/\text{dof} $). We enforce an upper limit of $ \chi^2/\text{dof} \leq 1.2 $ for our fits.
	
	As we will see, the $2s$ excitation sits at approximately 2.2 GeV presenting a
	formidable challenge to future calculations aiming to include all the multiparticle
	interpolating fields required below this energy. While our results will miss the subtle
	energy shifts associated with avoided level crossings of nearby scattering states,
	knowledge of these subtleties is not required to make progress in understanding the gross structure of $\Delta$-baryon resonances.
	
	\subsection{Simulation Details} \label{subsec:SimDets}
	Our calculations are based on the PACS-CS Collaboration's gauge-field ensembles \cite{PACS-CS:2008bkb}, made available via the ILDG. These configurations make use of the Iwasaki gauge action and an $ \mathcal{O}(a) $-improved Wilson quark action with Clover coefficient $ c_{\text{SW}} = 1.715 $, in full $ 2+1 $ flavour dynamical QCD. The simulations use $ \beta=1.90 $ on a $ 32^3\times 64 $ lattice.
	
	The masses of the up and down quarks are taken to be degenerate, and range over 5 values down to near the physical point. Meanwhile, the strange quark mass is held fixed. All masses are described by the Hopping parameter expansion with the Sommer parameter $ r_0 = 0.04921(64) $ fm used to set the physical scale, and the characteristics of each ensemble can be seen in Table~\ref{tab:LatSimParams}. The source position is held fixed at $ t_0 = 16 $, one-quarter of the lattice extent in the time direction. This allows for sufficient Euclidean time evolution of states to reduce excited state contamination, while reducing the impact of possible backwards propagating states reflected from the fixed boundary condition in the time direction. 
	
	For each mass ensemble, we apply a number of circular shifts to the gauge-fields, effectively allowing the source location to roam around the lattice $ N_\text{src} $ times. We then simply take an average over the resulting $ N_\text{src} $ sets of correlators for each configuration, allowing us to reduce the impact of the signal-to-noise ratio problem for baryons \cite{Parisi:1983ae,Lepage:1989hd,Hamber:1983vu,Detmold:2019ghl}. Jackknife uncertainties are calculated in a second-order single-configuration elimination calculation.
	
	\begin{table}
		\begin{center}
			\renewcommand{\arraystretch}{1.5}
			\begin{tabular}{cccccc}
				\hline
				\hline
				$ \kappa $ & $ m_\pi $ (MeV) & $ a $ (fm) & $ L $ (fm) & $ N_\text{cfgs} $ & $ N_\text{src} $ \\
				\hline
				0.13781 & 156 & 0.0933(13) & 2.9856(416) & 198 & 64 \\
				0.13770 & 293 & 0.0951(13) & 3.0432(416) & 400 & 16 \\
				0.13754 & 413 & 0.0961(13) & 3.0752(416) & 449 & 16 \\
				0.13727 & 572 & 0.1009(15) & 3.2288(480) & 397 & 32 \\
				0.13700 & 702 & 0.1023(15) & 3.2736(480) & 399 & 16 \\
				\hline
			\end{tabular}
			\caption{Table of parameters for our lattice ensembles, characterised by the Hopping parameter $ \kappa $.}
			\label{tab:LatSimParams}
		\end{center}
	\end{table}
	
	\section{Results: $ J^P = 3/2^+ $}\label{sec:Results1}
	We obtain results for each of the $ J^P $ combinations $ 1/2^+,\, 3/2^+,\, 1/2^-,\, 3/2^- $, by simply applying the corresponding pairs of spin and parity projection operators to the correlation functions of the correlation matrix.
	
	A wide range of variational parameters was considered including $t_0 = 17,\ 18,\ 19,\ 20,\ 21$
	and $\Delta t = 2,\ 3,\ 4$. While early values of $t_0$ and $\Delta t$ produce smaller
	uncertainties, better plateaus are obtained for larger values. However, large time
	considerations suffer from both exponential suppression of the excited-state information and the onset of large uncertainties. In general, we find that a choice of $ t_0 = 18 $ and $ \Delta t = 2 $ produces the most stable plateau fits across all 5 quark masses. This allows us to eliminate the variational parameters as having contributed any systematic uncertainty to the calculations as we compare across quark masses. We first present and discuss the results for the effective masses in the $ 3/2^+ $ channel, and follow this with identification of the radial excitations.
	
	\subsection{Spectrum} \label{subsec:MassSpec32+}
\begin{figure*}[]
		\includegraphics[width=0.8\linewidth]{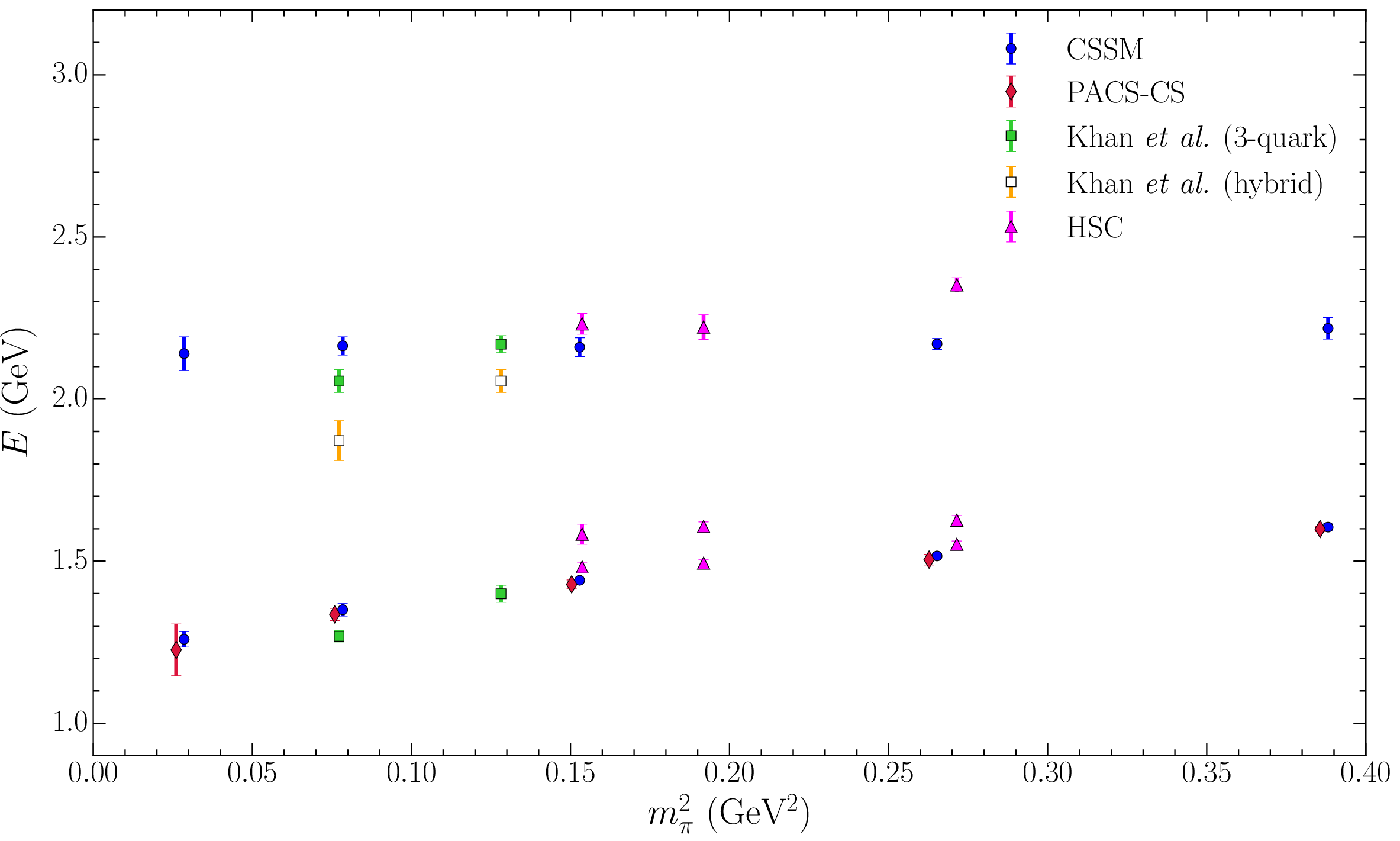}
		\caption{$ J^P = 3/2^+ \quad \Delta $ spectrum with CSSM lattice results from this investigation along with other comparable studies in the recent literature: PACS-CS \cite{PACS-CS:2008bkb}, HSC \cite{Bulava:2010yg}, and Khan {\it et al.} \cite{Khan:2020ahz}. In reporting results from Khan {\it et al.} we have differentiated states dominated by conventional three-quark operators and those having significant hybrid components. We omit the results for State 2, these are given in Table~\ref{tab:MassResults32+}.}
		\label{fig:DeltaLat32+}
	\end{figure*}

	The results of our computations are given in Table~\ref{tab:MassResults32+} as well as in Fig.~\ref{fig:DeltaLat32+} where data from other recent studies are shown for comparison \cite{PACS-CS:2008bkb,Khan:2020ahz,Bulava:2010yg}. Note that the studies labelled HSC and Khan {\it et al.} use different sets of interpolating fields and so are able to extract more states. We compare with the other studies to illustrate the systematic uncertainties associated with the calculations presented. All these simulation results producing large excitation energies necessarily do not include the multitude of multiparticle interpolatiing fields that	can participate at these energies. However, the three-quark interpolators considered are constructed using diverse techniques and agreement among the various collaborations brings credence to the results presented. Still, a variety of volumes are combined in these comparison plots and as such the avoided level crossings of the volume-dependent energy eigenstates will differ. These more subtle shifts are to be expected and can be used to inform the level of systematic uncertainty in the results presented. As our goal is to determine the approximate position of the $ 2s $ excitation of the $ \Delta(1232) $, the node information provided in this study is key.
	
	Starting with the lowest-lying state, we see that our results are consistent with the original PACS-CS results (offset slightly from ours in Fig.~\ref{fig:DeltaLat32+} for simplicity) and our lightest ensemble point improves on the original PACS-CS result in terms of the associated uncertainty. This can be attributed to our having averaged over $ 64 $ source locations as compared to their $ 4 $ source locations. The other lattice results shown in Fig.~\ref{fig:DeltaLat32+} reflect similar values of the ground state mass. As expected, there is a nonlinear dependence in the light quark mass region of low $ m_\pi^2 $ \cite{Leinweber:1999ig, Young:2002cj, Leinweber:2003dg}.
	
	Moving on to the first excitation results, we see resonable agreement between our results and the nearby data from other studies employing three-quark interpolators. We observe several features of note. Firstly, our excited state appears almost constant in pion mass, and secondly it lies well above the region that one would typically associate with the $ \Delta(1600) $. We do note however that the analysis of Khan {\it et al.} \cite{Khan:2020ahz} identifies lower-lying states with	a large hybrid operator component (highlighted in Fig.~\ref{fig:DeltaLat32+}) and therefore these are not associated with the three-quark excitation of our calculation. Their next states in the spectrum are dominated by three-quark operators and compare favourably with our results. In summary, it is reassuring that despite differences in methodology, lattice volumes and scale setting, our results are in approximate agreement with those of the other collaborations.
	
	The first low-lying excitations reported by the Hadron Spectrum Collaboration (HSC) \cite{Bulava:2010yg} next to their lowest-lying states are not observed in our analysis. A forthcoming analysis \cite{HockleyHEFTpaper} using Hamiltonian effective field theory to bring $\pi N$ scattering data in the $\Delta$-resonance channel to the finite-volume of the lattice indicates that the only finite-volume states in this energy region on a 2 fm lattice are the lowest-lying states. Thus, these states are not associated with the $\Delta(1600)$ but rather should be degenerate with the other lowest-lying HSC states presented. We defer further discussion on these particular HSC states to Ref.~\cite{HockleyHEFTpaper}.
	
	These considerations lead to a quite profound qualitative result; implementing smeared three-quark operators for describing the states in the $ \Delta(3/2^+) $ spectrum on the lattice appears insufficient for exciting states in the $ \Delta(1600) $ resonance region. The question arises then; is the ground state a simple $ 1s $ state as one expects, and is the first excitation that we have calculated near $ \sim 2.15 $ GeV the corresponding $ 2s $ excitation? This would give some tension with the long-held stance that the $ \Delta(1600) $ is the radial excitation of the $ \Delta(1232) $. The identification of radial excitations is explored below.

	\begin{table*}[]
		\begin{center}
			\renewcommand{\arraystretch}{1.25}
			\setlength\tabcolsep{4mm}
			\begin{tabular}{c|cc|cc|cc}
				\hline
				\hline
				$ m_\pi $ (MeV) & $ m_0 $ (GeV) & $ \Delta m_0 $ (GeV) & $ m_1 $ (GeV) & $ \Delta m_1 $ (GeV) & $ m_2 $ (GeV) & $ \Delta m_2 $ (GeV) \\
				\hline
				156 & 1.261 & 0.023 & 2.140 & 0.052 & 3.097 & 0.172 \\
				293 & 1.350 & 0.019 & 2.164 & 0.028 & 2.967 & 0.051 \\
				413 & 1.441 & 0.006 & 2.160 & 0.029 & 2.871 & 0.033 \\
				572 & 1.516 & 0.006 & 2.170 & 0.017 & 2.808 & 0.026 \\
				702 & 1.605 & 0.010 & 2.218 & 0.033 & 2.859 & 0.041 \\
				\hline
			\end{tabular}
			\caption{Masses for the $ 3/2^+ $ $ \Delta $ spectrum.}
			\label{tab:MassResults32+}
		\end{center}
	\end{table*}
		
	\subsection{Radial Excitations}\label{ssRadEx}
	With the lattice masses calculated, we now identify these in terms of radial excitations. Our approach to this is based on counting the number of nodes in the wave function, as is discussed in detail in \cite{Mahbub:2013ala, Roberts:2013oea, Roberts:2013ipa}. We present here a brief summary of the techniques involved before discussing our results.
	
	In solving the GEVP of the variational method, we build up a description of the wave function for each state in the spectrum as a superposition of Gaussian distributions. An important feature of this is the possible presence of nodes in the wave function.
	
	Consider, for example, the case of a narrow Gaussian of positive signature, superposed with a broader Gaussian of negative signature. As we consider points further away from the centre of the distribution, the positive Gaussian will diminish and the negative Gaussian will begin to dominate the superposition. As a result, there will be some radius at which the superposition crosses through zero. The associated probability amplitude will thus possess a single radial node, corresponding to a $ 2s $ state. For a superposition with additional Gaussians which alternate in signature as the widths increase, one can produce more nodes in the wave function.
	
	These techniques of superposing smeared quark sources and sinks through a correlation
	matrix analysis were used to study the wave functions of even-parity nucleon excitations
	in Refs.~\cite{Roberts:2013oea,Roberts:2013ipa}. There, the linear combination of smeared sources defined in the GEVP analysis was used at the source to excite the energy eigenstates. After fixing the gauge fields to Landau gauge, the quark fields in the sink operator were given individual spatial positions such that they could be pulled apart. For example, fixing the two $u$ quarks of the proton at the centre of the lattice, the wave function of the $d$-quark was resolved by studying the strength of the baryon annihilation matrix element as a function of the $d$-quark position. Upon observing nodes in the $d$-quark wave function, it was noted that this information was already apparent in the correlation matrix eigenvectors for the energy eigenstates, manifested by sign alternating eigenvector components as the Gaussian widths of the sources increase. It is this method that is used here to identify nodes.
	
\begin{figure*}[]
		\includegraphics[width=0.8\linewidth]{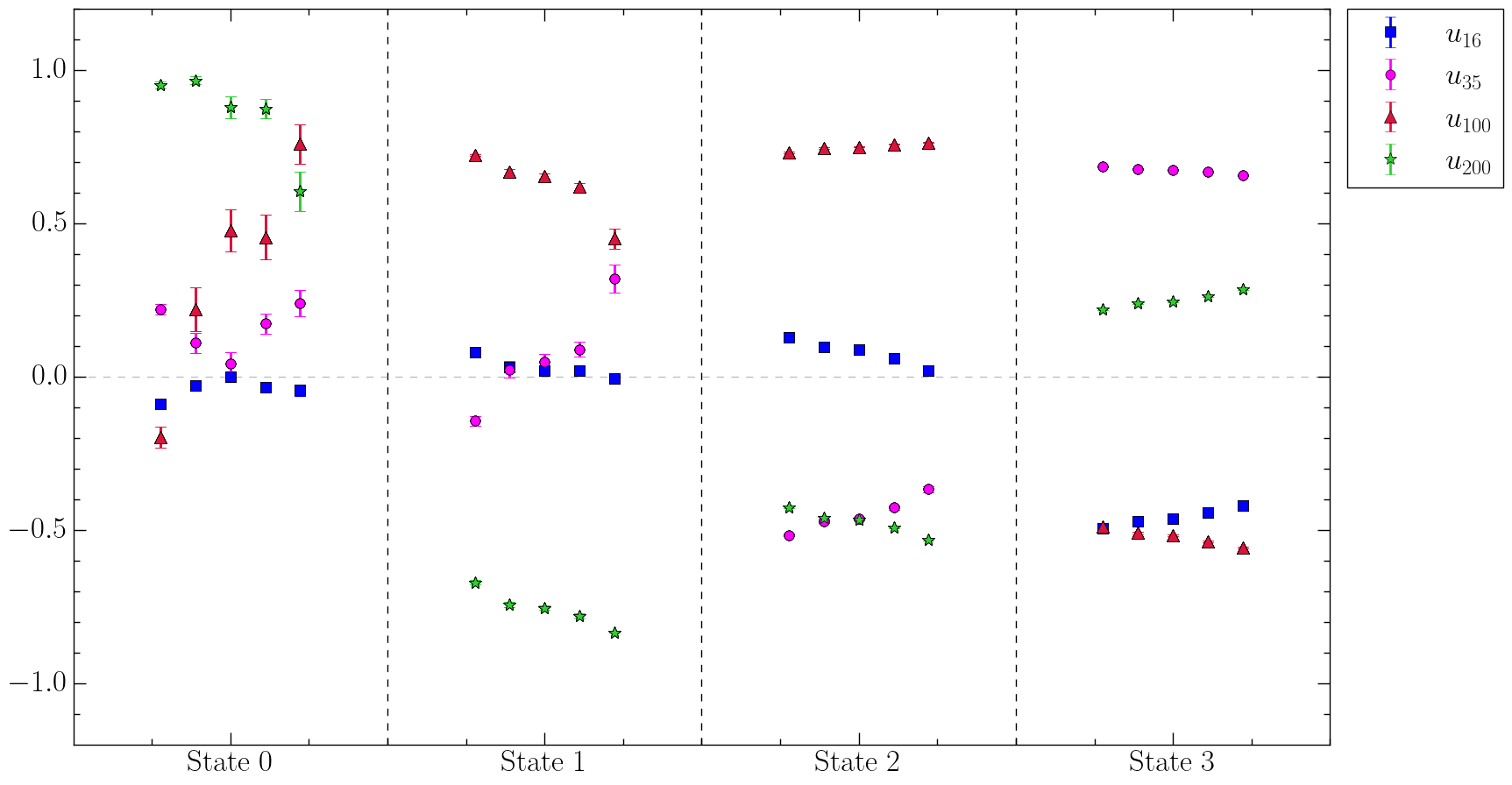}
		\caption{Plot of the $ \boldsymbol{u} $ eigenvector components for each of the states in our $ 4\times4 $ correlation matrix analysis. For each state, the quark mass increases from left to right, and the different smeared Gaussian components are denoted by the various coloured symbols. Noting the small values near zero introduce ripples but do not change the sign of the wave function within the lattice volume, one can infer the presence of nodes by counting the number of crossings through zero at a particular quark mass as the smearing number increases from 16 to 200. We find state $ \alpha $ has $ \alpha $ nodes when counting from zero.}
		\label{fig:uvec_plot}
	\end{figure*}

	We can view our GEVP procedure as constructing the superposition of Gaussians which best describes the wave function of the states in the spectrum, that is, solving the coefficients $ u^\alpha_i $ in
	\begin{equation}
		\phi^\alpha = u^\alpha_{16}\, \chi_{16} + u^\alpha_{35}\, \chi_{35} + u^\alpha_{100}\, \chi_{100} + u^\alpha_{200}\, \chi_{200}
	\end{equation}
	such that $ \phi^\alpha $ is a good approximation to the wave function for energy level $ \alpha $. The values for the various $ u^\alpha_i $ are displayed in Fig.~\ref{fig:uvec_plot}.
	
	To reiterate then, we start by considering the $ u_{16} $ components (the weights for the narrowest Gaussian) and observe whether the weight changes sign as we move on to the $ u_{35} $. If it does, we would associate this with a node. We repeat this process as we continue onto the $ u_{100} $ and $ u_{200} $. 
			
	For the eigenvector components to have a physical interpretation, a normalisation
	prescription is required for the baryon interpolating fields of various smearing levels. We
	proceed by introducing a real scaling factor, $s_i$, for each baryon interpolating field, $\chi_i	\to s_i\, \chi_i$, and normalising the diagonal spin-parity projected correlation functions of Eq.~(\ref{eq:trace}) to one at one time slice after the source, {\it i.e.} $G_{ii}(t-t_0=1,\, \boldsymbol{p}=0) = 1$ for each $i=16,\,35,\,100,\,200$. This normalisation condition for the diagonal correlators having the same smeared	source and sink provides a constraint for the determination of the scaling factor $s_i$. With	the $\chi_i$ suitably normalised, the coefficients $u_i^\alpha$ may then be applied to normalised Gaussians to gain insight into the nature of the interpolating field wave function.
	
\begin{figure*}[]
		\includegraphics[width=0.33\linewidth]{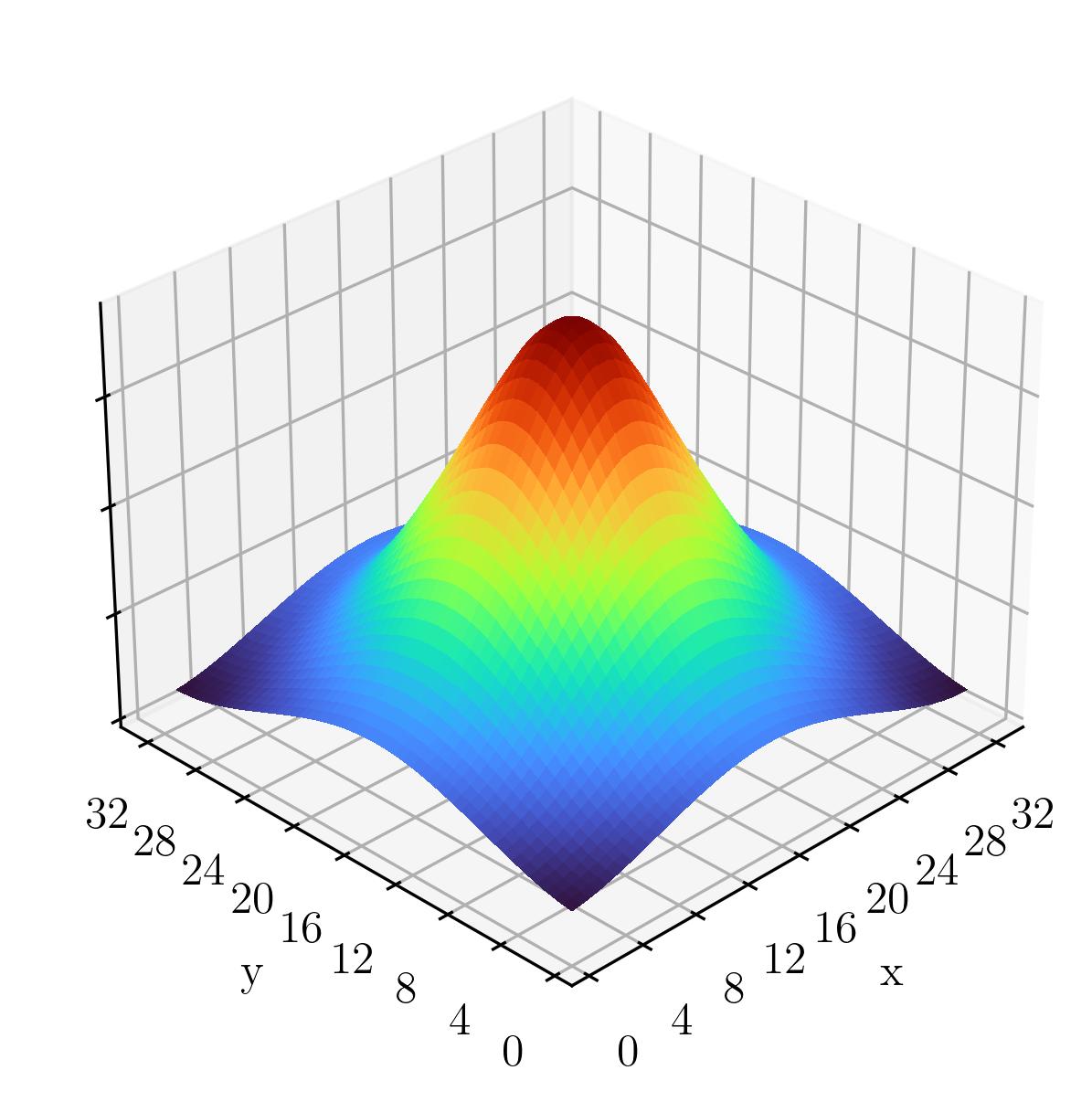}\hfil
		\includegraphics[width=0.33\linewidth]{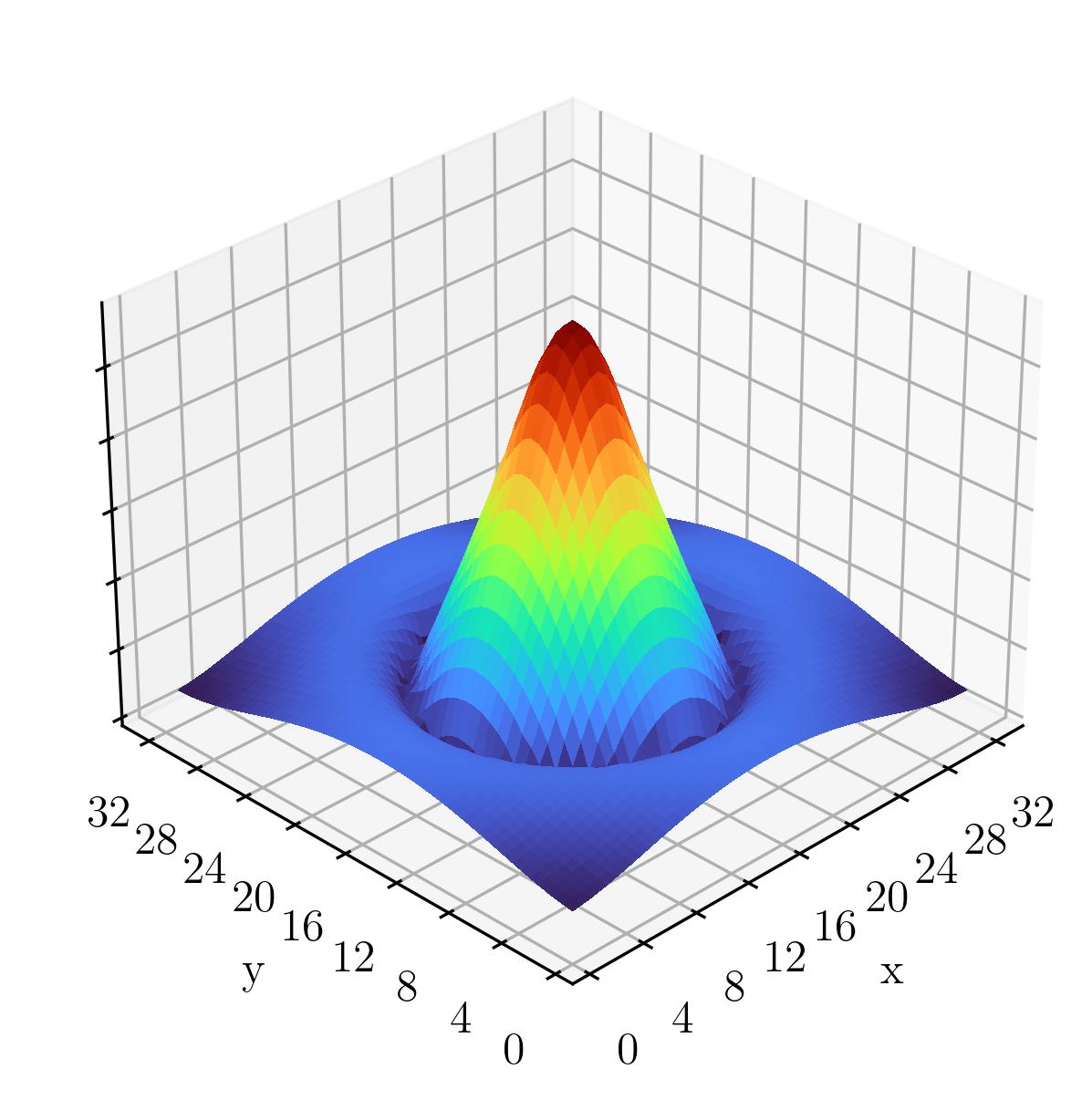}\hfil
		\includegraphics[width=0.33\linewidth]{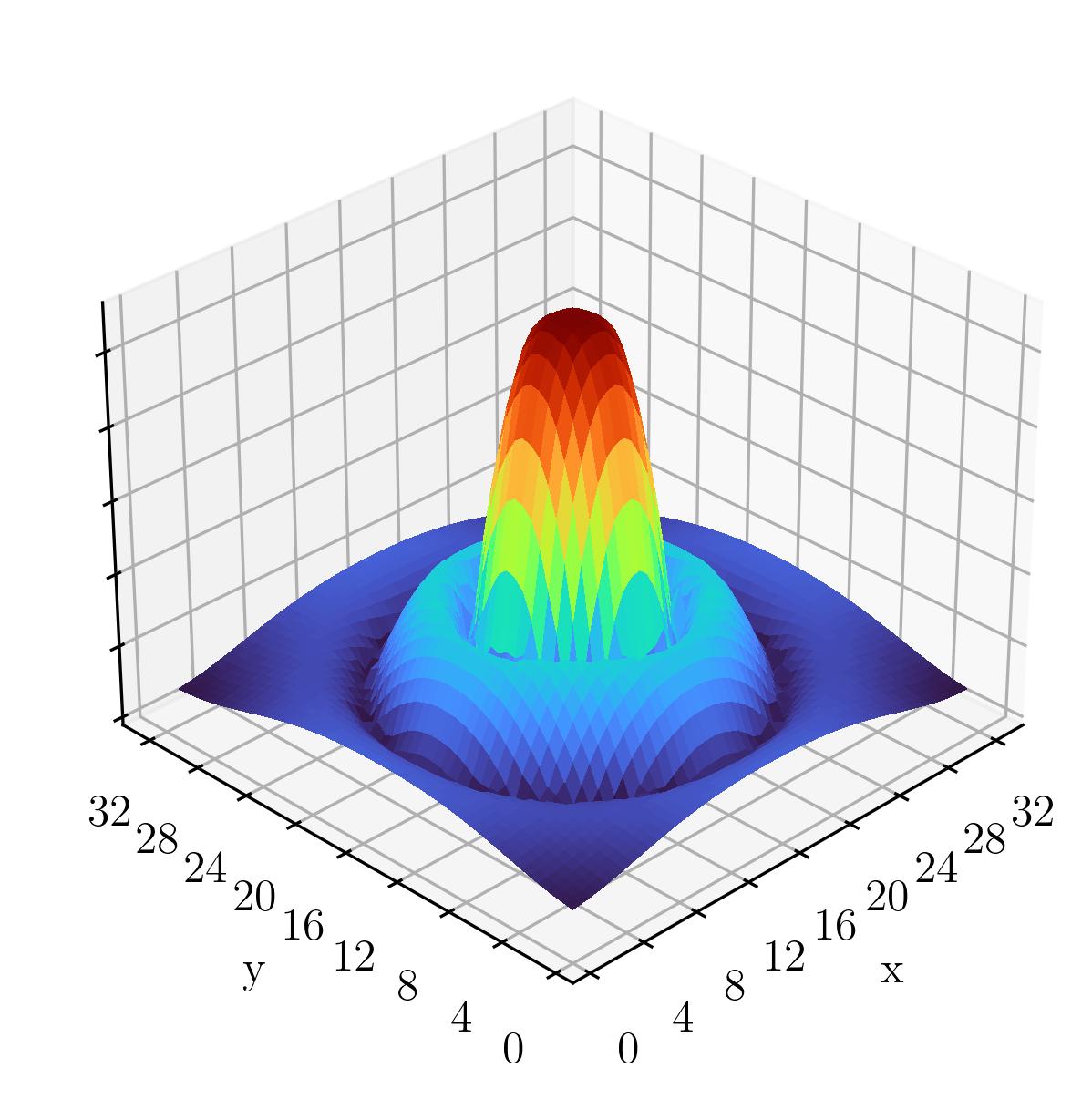}
		\caption{Probability amplitude distributions for the $ 3/2^+ $ $ \Delta $ spectrum, constructed from the superposition of smeared interpolating fields, based on the middle quark mass ensemble. These are obtained by taking a superposition of normalised smeared sources with weights given by the $ \boldsymbol{u} $ vector components in Fig.~\ref{fig:uvec_plot}. Nodes in the wave function are indicated by radially symmetric regions of dark blue away from the edges of the volume. (Left) No nodes are present in State 0, a $ 1s $ state. (Center) A single radially symmetric node is present in State 1, so this is identified as a $ 2s $ state. (Right) an inner and outer node are visible in State 2, a $ 3s $ state.}
		\label{fig:wavefun_plots}
	\end{figure*}

	We can visualise these superpositions by simply adding up normalised Gaussian distributions weighted by $ u^\alpha_i $ for each state $ \alpha $. Taking a $ 2 $-dimensional slice of our lattice in the $ x-y $ plane, the square of this superposition can then be plotted as a probability amplitude as shown in Fig.~\ref{fig:wavefun_plots} for the $ m_\pi = 413 $ MeV ensemble. Indeed, one can see the presence of nodes in both Figs.~\ref{fig:uvec_plot} and \ref{fig:wavefun_plots}. 
		
	Consider Fig.~\ref{fig:uvec_plot} first, and in particular the middle quark mass results for the $ \boldsymbol{u} $ vector components. Small eigenvector components will introduce ripples in the wave functions but are not large enough to change the sign within the lattice volume. In this light, State 0 has all of its large components being non-negative, so there is no node visible in the probability amplitude illustrated in the left-hand plot of Fig.~\ref{fig:wavefun_plots}. In the case of State 1, we start with $ u_{16} \sim u_{35} \sim 0 $ and then the remaining components are both far from zero, with $ u_{100} > 0 $ and $ u_{200} < 0 $. This case is effectively the same as the toy example above where a broad Gaussian is being subtracted from a narrow Gaussian. The result is that we expect a single node in the probability amplitude and this is reflected in the right-hand plot of Fig.~\ref{fig:wavefun_plots}. Turning our attention to State 2 in Fig.~\ref{fig:uvec_plot}, we note the $u_{16}$ values are small relative to the $u_{35}$ values and do not create a node. However sign oscillations from $u_{35}$ to $u_{100}$ to $u_{200}$, all of which have significant magnitudes, induce 2 nodes in the interpolator wave function. Similarly, State 3 has 3 nodes.
	
	At this point we have identified the masses of three states in the spectrum, and have further classified these as being a $ 1s,\, 2s $ and $ 3s $ state. Interestingly, while the $ 1s $ state aligns well with the $ \Delta(1232) $, the $ 2s $ state has a mass around $ \sim 2.15 $ GeV so cannot be associated in a significant manner with the $ \Delta(1600)$. Thus it appears that this state is associated with 2- and 3-body scattering channels. The use of Hamiltonian Effective Field Theory (HEFT) as a L\"{u}scher adjacent method will allow us to provide insight in a forthcoming paper into which channels may be useful in future lattice studies for examining the $ \Delta(1600) $ resonance.
	
	\section{Additional Results: $ J^P = 3/2^-, 1/2^\pm $} \label{sec:Results2}
\begin{figure*}
		\includegraphics[width=0.9\linewidth]{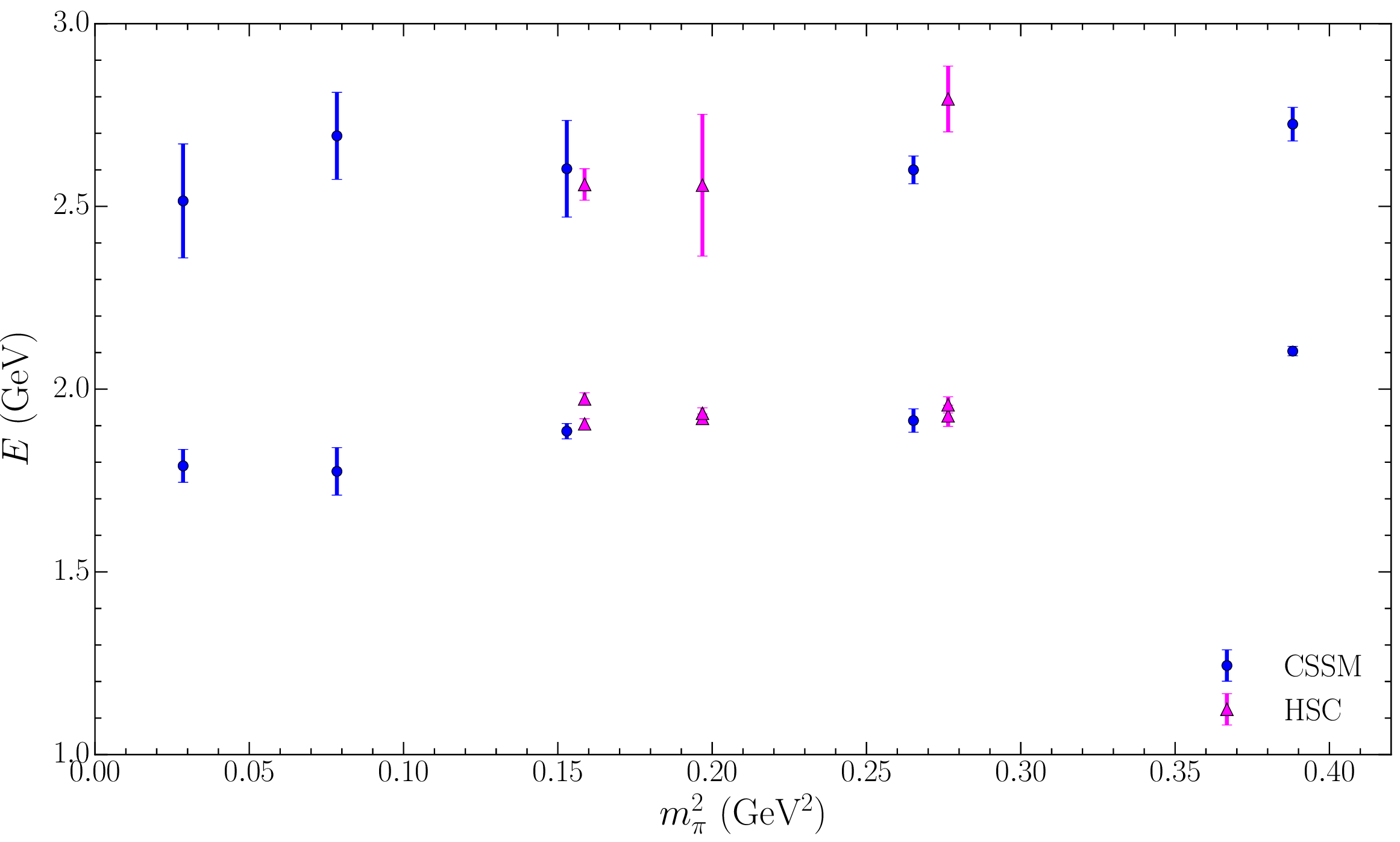}
		\caption{$ J^P = 3/2^- \ \Delta $ spectrum. CSSM results from this study are compared with results from the HSC \cite{Bulava:2010yg}. Note that the HSC results include 3 states at each pion mass.}
		\label{fig:DeltaLat32-}
	\end{figure*}

\begin{figure*}
		\includegraphics[width=0.9\linewidth]{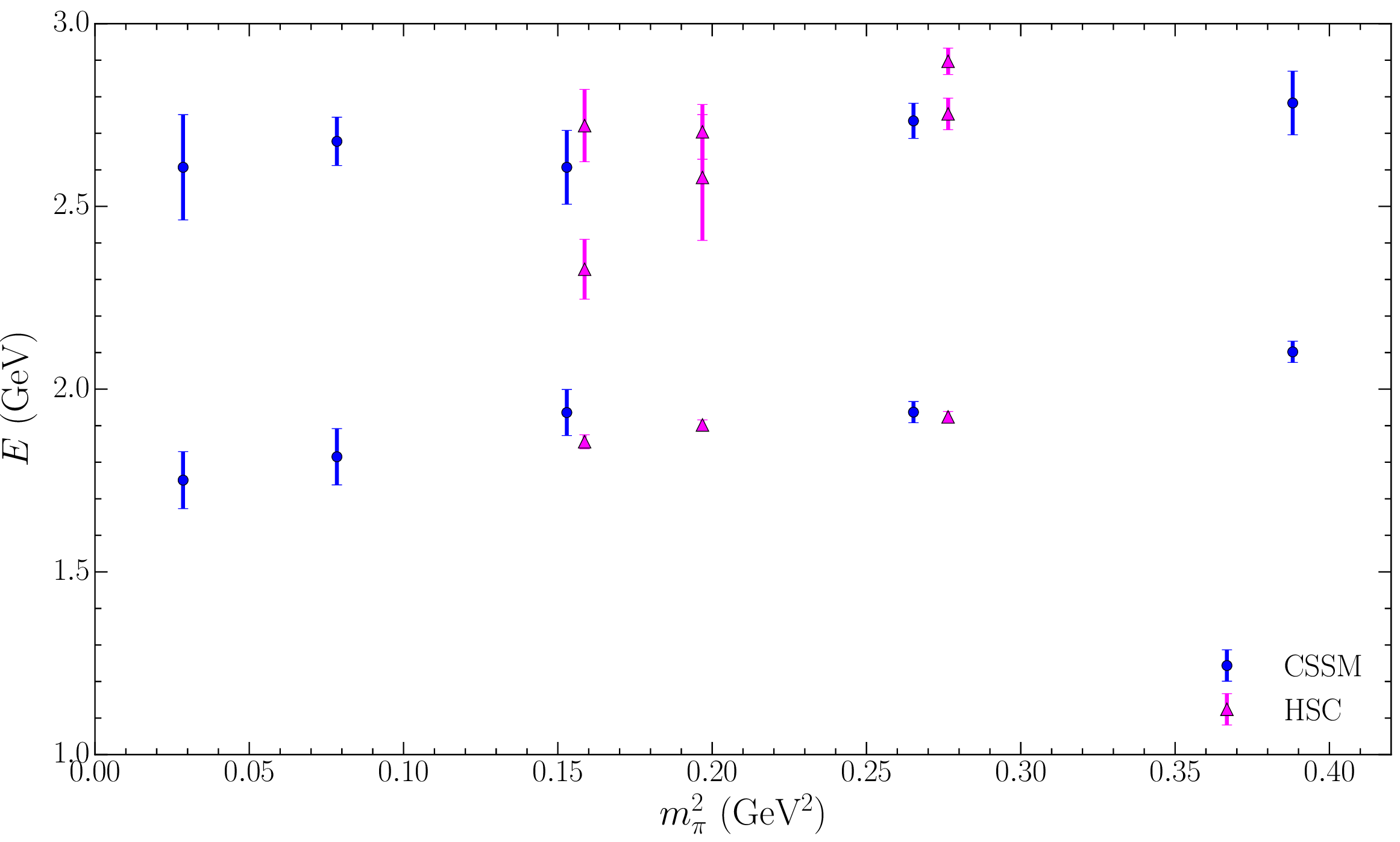}
		\caption{$ J^P = 1/2^- \ \Delta $ spectrum. CSSM results from this study are compared with results from the HSC \cite{Bulava:2010yg}. }
		\label{fig:DeltaLat12-}
	\end{figure*}

\begin{figure*}
		\includegraphics[width=0.9\linewidth]{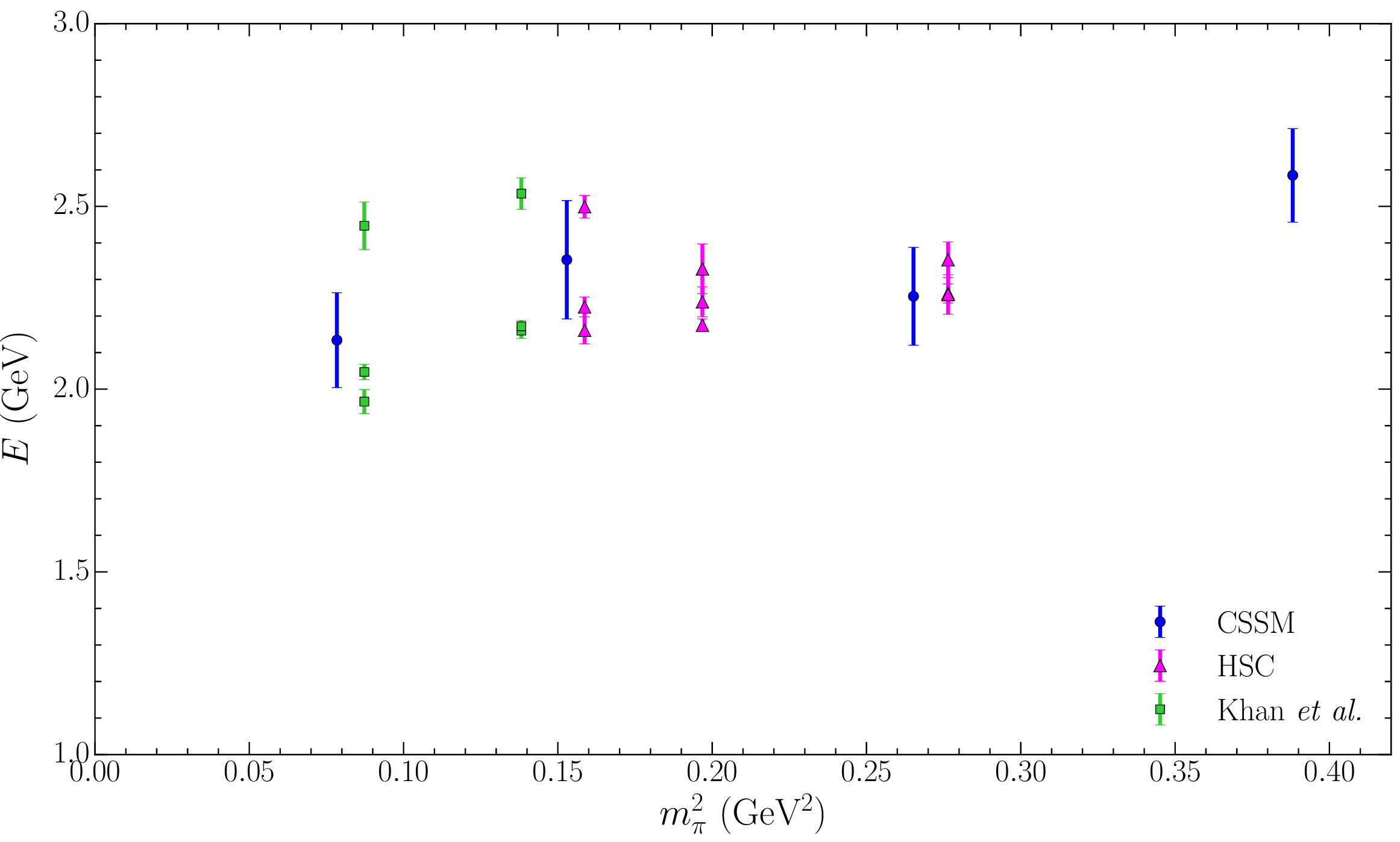}
		\caption{$ J^P = 1/2^+ \ \Delta $ spectrum. CSSM results from this study are compared with the HSC \cite{Bulava:2010yg} and Khan {\it et al.} \cite{Khan:2020ahz}. Note that the HSC and Khan {\it et al.} results include 3 states at each pion mass.}
		\label{fig:DeltaLat12+}
	\end{figure*}

	While the spin-3/2 even-parity case is of most interest in the current study, we also report results for the spin-1/2 states as well as odd-parity states for both spins. The spectra for these cases are given in Figs.~\ref{fig:DeltaLat32-}, \ref{fig:DeltaLat12-} and \ref{fig:DeltaLat12+}, with numerical values for our fits given in Tables~\ref{tab:MassResults32-}, \ref{tab:MassResults12-} and \ref{tab:MassResults12+}. While we are not able to extract excited states for all of these cases, we can observe a distinct trend in the ground state masses: 
	\begin{itemize}
		\item the $ 3/2^+ $ has the lowest mass.
		\item the $ 3/2^- $ and $ 1/2^- $ are a little higher in energy, with similar masses.
		\item the $ 1/2^+ $ has a mass well above these states.
	\end{itemize}
	
	This is consistent with previous CSSM results using quenched QCD \cite{Zanotti:2003fx}, and has good alignment with the resonances observed in experiment \cite{ParticleDataGroup:2022pth}.
			
	\begin{table*}[]
		\begin{center}
			\renewcommand{\arraystretch}{1.25}
			\setlength\tabcolsep{4mm}
			\begin{tabular}{c|cc|cc|cc}
				\hline
				\hline
				$ m_\pi $ (MeV) & $ m_0 $ (GeV) & $ \Delta m_0 $ (GeV) & $ m_1 $ (GeV) & $ \Delta m_1 $ (GeV) & $ m_2 $ (GeV) & $ \Delta m_2 $ (GeV) \\
				\hline
				156 & 1.790 & 0.045 & 2.515 & 0.156 & 3.408 & 0.126 \\
				293 & 1.775 & 0.065 & 2.693 & 0.119 & 3.374 & 0.111 \\
				413 & 1.885 & 0.021 & 2.603 & 0.132 & 3.182 & 0.055 \\ 
				572 & 1.914 & 0.032 & 2.600 & 0.038 & 3.182 & 0.042 \\
				702 & 2.104 & 0.013 & 2.725 & 0.046 & 3.310 & 0.060 \\
				\hline
			\end{tabular}
			\caption{Masses for the $ 3/2^- $ $ \Delta $ spectrum.}
			\label{tab:MassResults32-}
		\end{center}
	\end{table*}
			
	\begin{table*}[]
		\begin{center}
			\renewcommand{\arraystretch}{1.25}
			\setlength\tabcolsep{4mm}
			\begin{tabular}{c|cc|cc}
				\hline
				\hline
				$ m_\pi $ (MeV) & $ m_0 $ (GeV) & $ \Delta m_0 $ (GeV) & $ m_1 $ (GeV) & $ \Delta m_1 $ (GeV) \\
				\hline
				156 & 1.751 & 0.078 & 2.607 & 0.144  \\
				293 & 1.815 & 0.077 & 2.678 & 0.066  \\
				413 & 1.936 & 0.063 & 2.607 & 0.101  \\
				572 & 1.937 & 0.029 & 2.734 & 0.048  \\
				702 & 2.102 & 0.029 & 2.783 & 0.087  \\
				\hline
			\end{tabular}
			\caption{Masses for the $ 1/2^- $ $ \Delta $ spectrum.}
			\label{tab:MassResults12-}
		\end{center}
	\end{table*}	
		
	\begin{table*}[]
		\begin{center}
			\renewcommand{\arraystretch}{1.25}
			\setlength\tabcolsep{4mm}
			\begin{tabular}{c|cc}
				\hline
				\hline
				$ m_\pi $ (MeV) & $ m_0 $ (GeV) & $ \Delta m_0 $ (GeV) \\
				\hline
				293 & 2.134 & 0.130 \\
				413 & 2.354 & 0.162 \\
				572 & 2.254 & 0.134 \\
				702 & 2.585 & 0.128 \\
				\hline
			\end{tabular}
			\caption{Masses for the $ 1/2^+ $ $ \Delta $ spectrum. Note that we are only able to identify plateaus in the effective mass series for the lowest-lying state at the 4 heavier quark masses.}
			\label{tab:MassResults12+}
		\end{center}
	\end{table*}
	
	\subsection{Spectra}
	Considering the resonance positions in Table~\ref{tab:DeltaPDG}, we note alignment between our results and those from the PDG, in that the ground state $ 1/2^- $ and $ 3/2^- $ states are nearby to one another, and likewise for the corresponding excited states. These appear to correspond approximately to the $ \Delta(1620)(1/2^-) $ and $ \Delta(1700)(3/2^-) $, and the $ \Delta(1900)(1/2^-) $ and $ \Delta(1940)(3/2^-) $, respectively. This suggests these resonances may be single-particle dominated states dressed by their decay channels.
	
	Furthermore, there is excellent agreement across all our lowest-lying masses with those reported by HSC, as well as Khan {\it et al}. Our $ 3/2^- $ results are competitive with those of the other studies, and show fair agreement even at the first excitation level. 
	
	The $ 1/2^+ $ case presents the greatest challenge to our approach. Even the lowest-lying state could not be extracted at the lightest pion mass, and no excited states were able to be extracted at any pion mass. This is due to the initial interpolating field we use being constructed for coupling primarily to spin-3/2 states. Once we project both the spin and parity, there is little signal left for the spin-1/2 case. This is particularly problematic in the $ 1/2^+ $ case since the lowest-lying state is expected to lie much higher in energy than the other cases.
	
	While the $ 1/2^+ $ results are clearly not competitive with those of the other groups, there is approximate agreement with their findings. A more suitable choice of interpolator basis is required for further study of this channel. 
			
	\subsection{Radial Excitations}
	Similarly to the $ 3/2^+ $ results, we can also consider the $ \boldsymbol{u} $ vector components for each of these states, and visualise the resulting descriptions of the wave functions. However, there is a subtle point worth mentioning here, specifically how the odd-parity properties are accessed. In particular, as was outlined in Section~\ref{subsec:BarSpec}, the parity nature of our correlators is dictated by the presence of a $ \gamma_5 $ matrix in the overlap of interpolating fields and baryon states. We thus expect that odd-parity character should be contained in only the spinor part of the wave function. This means that even for the odd-parity baryons, we should expect a radially symmetric description of the spatial wave function, just as was obtained in Fig.~\ref{fig:wavefun_plots}.

\begin{figure*}[]
		\includegraphics[width=0.3\linewidth]{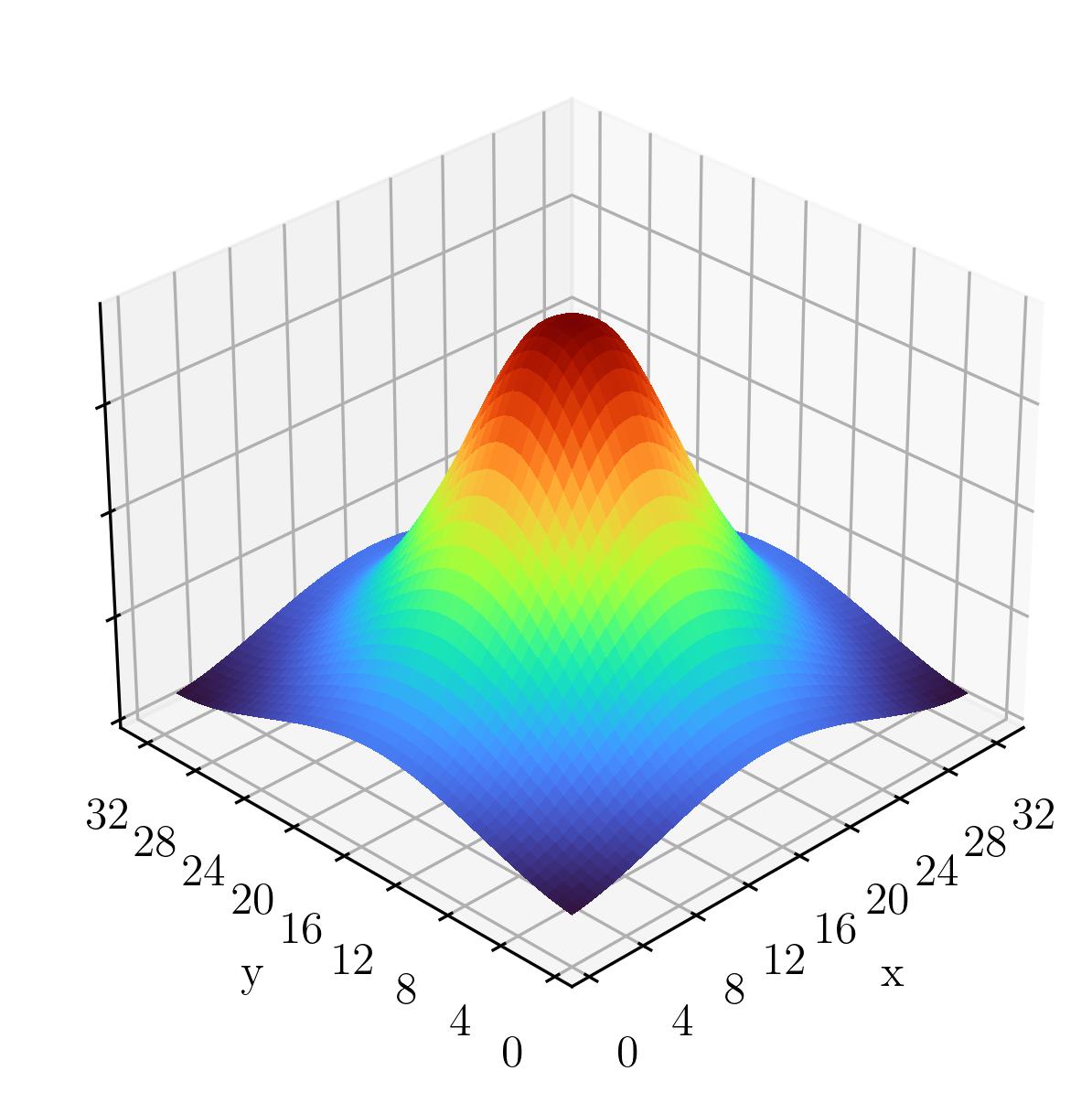}
		\includegraphics[width=0.3\linewidth]{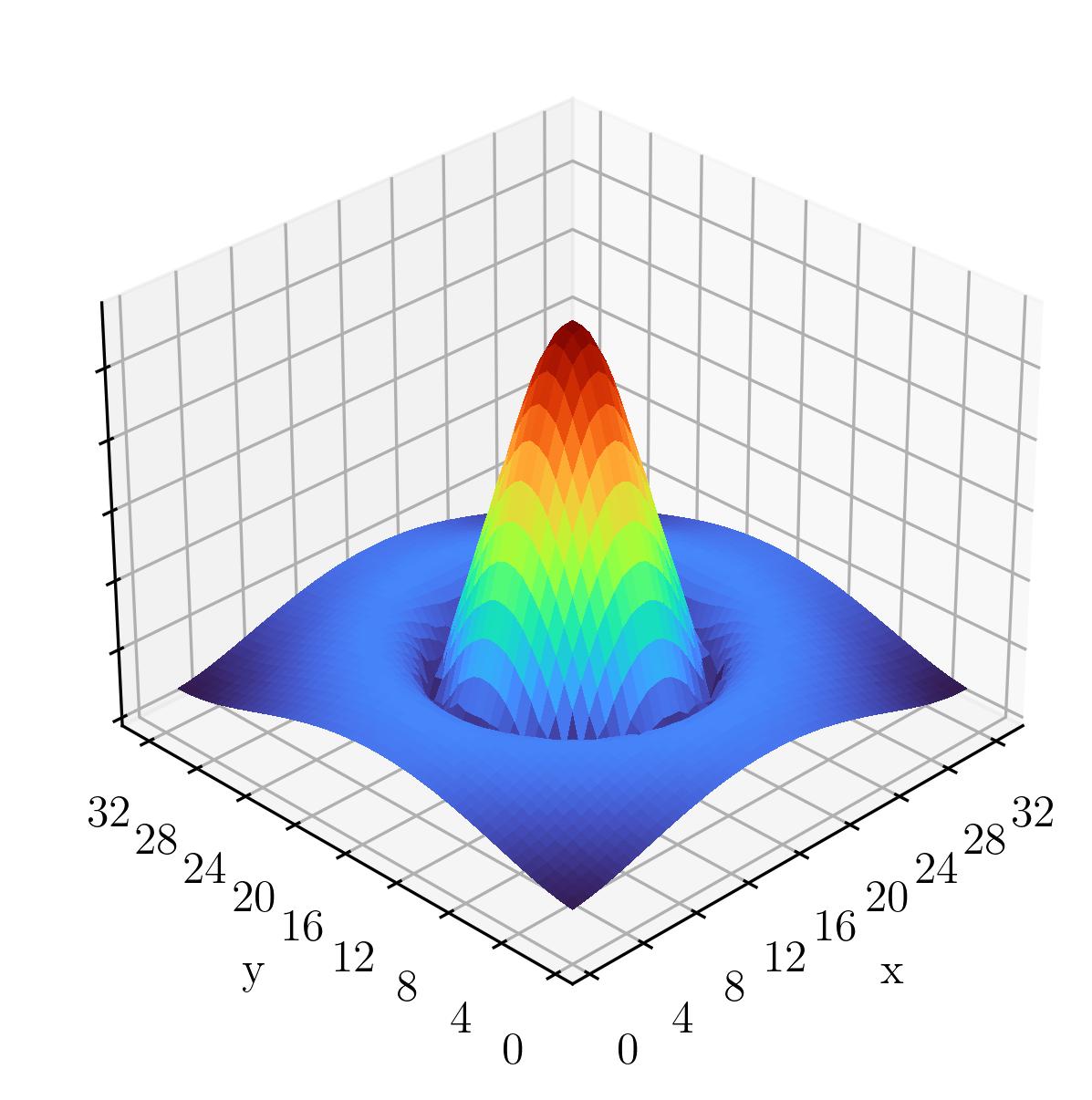}
		\includegraphics[width=0.3\linewidth]{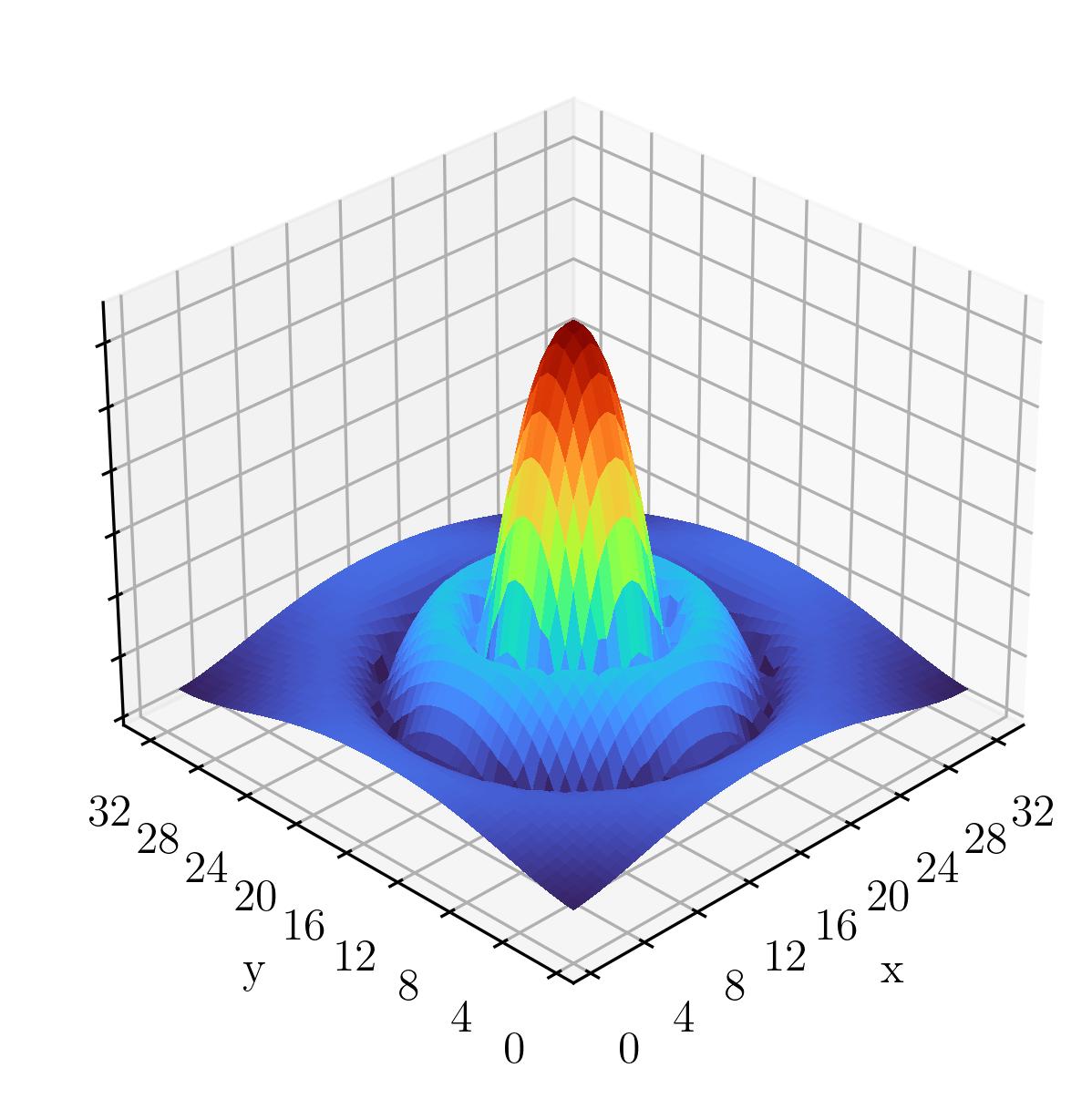}
		\caption{$ 3/2^- $ probability amplitudes, based on the superposition of smeared interpolating fields on the middle quark mass ensemble. These symmetric wave functions multiply  the lower components of the Dirac spinors to generate odd-parity states. Nodes in the wave function are indicated by radially symmetric regions of dark blue away from the edges of the volume.}
		\label{fig:wavefn32-}
	\end{figure*}

\begin{figure*}[]
		\includegraphics[width=0.4\linewidth]{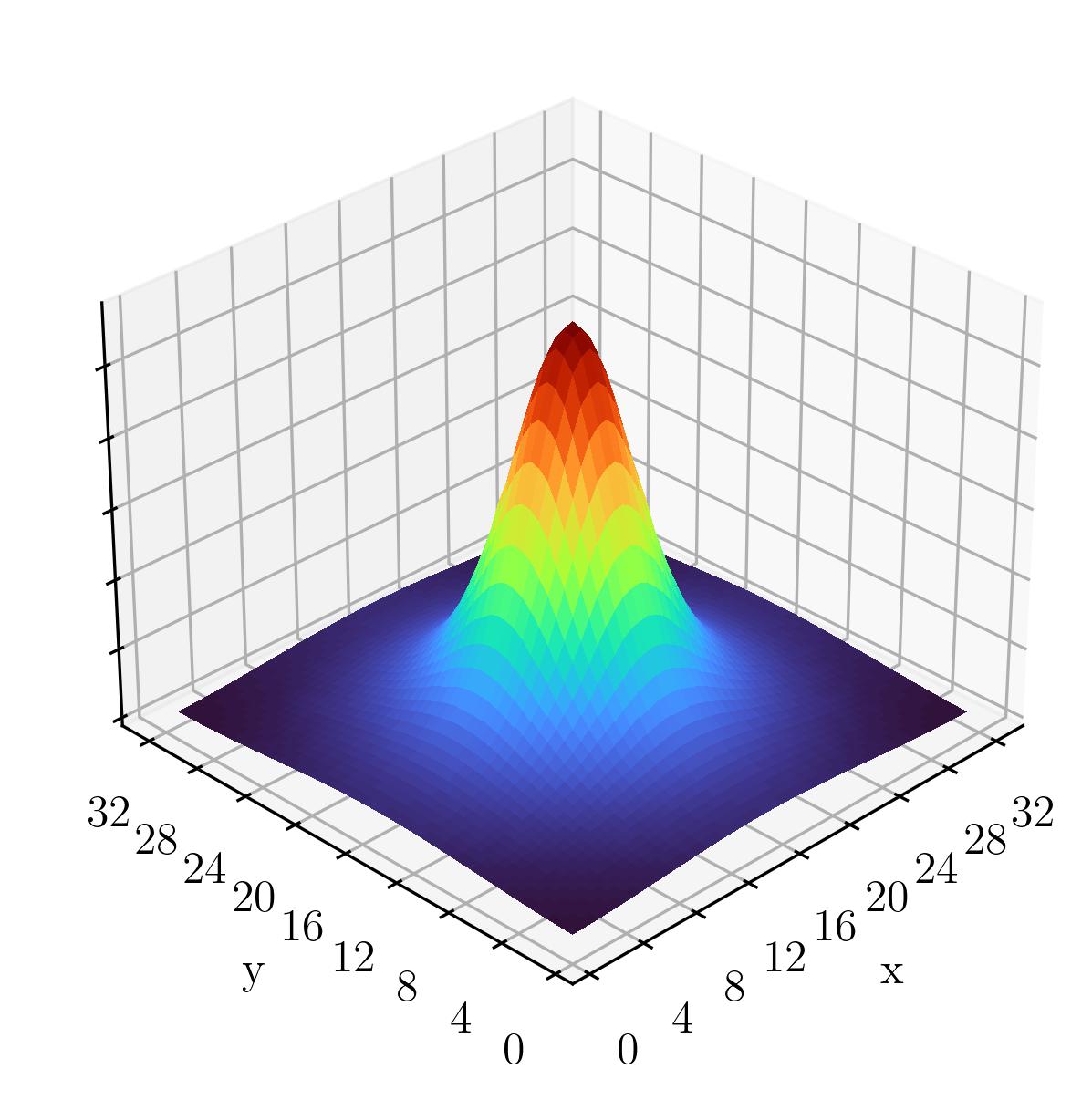}
		\includegraphics[width=0.4\linewidth]{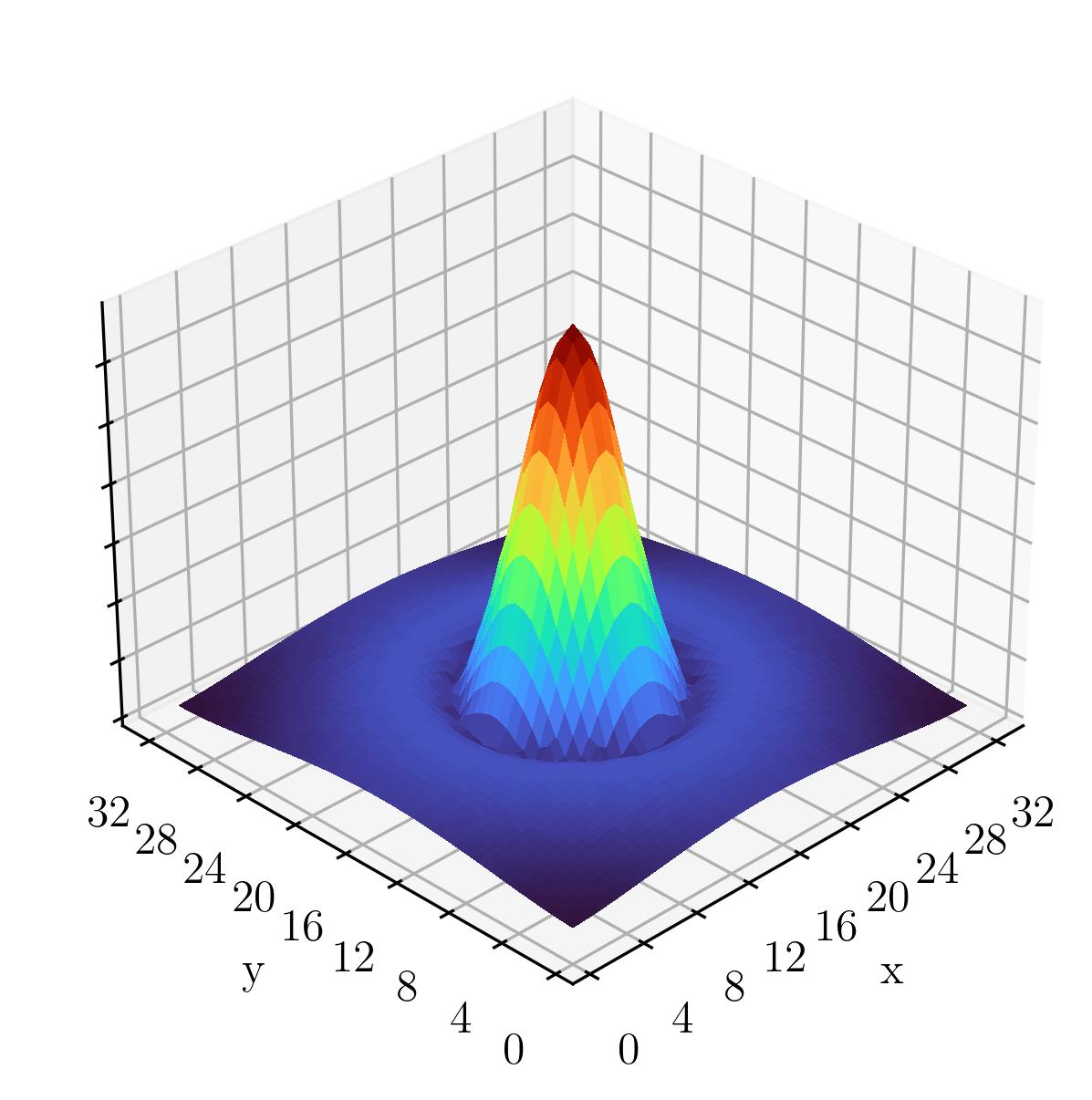}
		\caption{$ 1/2^- $ probability amplitudes, based on the superposition of smeared interpolating fields on the middle quark mass ensemble. These symmetric wave functions multiply  the lower components of the Dirac spinors to generate odd-parity states. Nodes in the wave function are indicated by radially symmetric regions of dark blue away from the edges of the volume.}
		\label{fig:wavefn12-}
	\end{figure*}

	The visualisations of the interpolator wave functions for the $ 3/2^- $ and $ 1/2^- $ states are shown in Figs.~\ref{fig:wavefn32-} and \ref{fig:wavefn12-}. Note that while we are able to cleanly extract all $ \boldsymbol{u} $ vector components for all 4 energy states in each spin-parity pairing, we only show the wavef function plots for which we were also able to extract masses. The $ 1/2^+ $ case can also be assigned a Gaussian superposition for each energy state, but since we were only able to extract a ground state energy in this case, we omit the plots of the associated Gaussian distributions.

	Once again, we see that the first excited state involves a node in its structure. For the $ 3/2^- $ case, our results suggest that the $ \Delta(1940)(3/2^-) $ may not be a simple excitation of the $ \Delta(1700)(3/2^-) $, since it lies well below the excitation energy determined herein. This resonance may be associated with one or more scattering channels in a non-trivial way, as has been suggested for the $ \Delta(1600)(3/2^+) $ \cite{Abell:2021awi}.
	
	For the $ 1/2^- $ case, there is doubt whether the $ \Delta(1900)(1/2^-) $ or even the $ \Delta(2150)(1/2^-) $ are simple radial excitations of the $ \Delta(1620)(1/2^-) $, owing to a substantial energy gap between these and the observed $ 2s $ excitation energy on the lattice. Again, a coupled-channel analysis approach may shed some light on the structure of these resonances.
		
\section{Conclusion}\label{sec:Conclusion}
We have presented an analysis of the low-lying $ \Delta $-baryons using techniques within the framework of lattice QCD. Our focus is on identifying the position of the $ 2s $ excitation of the $ \Delta(1232)(3/2^+) $ in the spectrum. Our technique of superposing smeared fermion fields in constructing our baryon sources lends itself to identifying the nodes of wave functions, thereby identifying radial excitations of states obtained by the variational method. This has allowed us to conclude the following for the energy eigenstates observed in our calculations:
\begin{enumerate}
	\item The lowest-lying state having an energy in the $ \Delta(1232)(3/2^+) $ resonance region is identified as a $ 1s $ state.
	\item The first excitation has an energy of $ 2.14(5) $ GeV in the lightest mass ensemble, with only weak dependence on quark mass evident across the considered ensembles. This state is identified as a $ 2s $ state, and thus the radial excitation of the $ \Delta(1232)(3/2^+) $. There is a substantial energy gap between this state and the $ \Delta(1600)(3/2^+) $ and it is unlikely to be strongly associated with this resonance. We note that this situation is similar to that observed in Refs.~\cite{Roberts:2013oea,Liu:2016uzk} where the $2s$ excitation was resolved at 1.90(6) GeV, well above the position of the Roper
	resonance at 1.44 GeV.
	\item The second excitation is identified as a $3s$ state with two nodes in the wave function.
	The energy of this state on our 3 fm lattice is large at 3.10(17) GeV. When compared with
	the $1s$ and $2s$ energies of 1.26(2) and 2.14(5) GeV respectively, one observes a
	mass splitting of $\sim0.9$ GeV between the radial excitations, very large compared to
	conventional quark models.	
\end{enumerate}

We conclude that the missing baryon resonances problem of the last three decades
follows from incorrectly tuning quark model parameters to reproduce the $N(1440)$ Roper
resonance and the $\Delta(1600)$ resonance as $2s$ radial excitations. We now know
the $2s$ radial excitations sit at 1.9 and 2.1 GeV for the $N$ and $\Delta$ respectively.
Thus there is an important opportunity for the development of modern quark models
whose spectra are compatible with this modern understanding of quantum field theory.

Aside from the main results for the $ 3/2^+ $ channel, we have also reported results for the $ 3/2^-,\, 1/2^+ $ and $ 1/2^- $ spectra. Here we resolved a node in the first excitations of the $1/2^-$ and $3/2^-$states illustrating the internal structure of these states for the first time. Our spectra are qualitatively similar to other collaborations pursuing these high-energy excitations.

Since a simple three-quark dominated model appears to be insufficient to describe the structure of the $ \Delta(1600) $, attention must turn to interactions in multi-particle meson-baryon channels. It is expected that the dominant scattering channels of $ \pi N $ and $ \pi \Delta $ will play a key role in describing this resonance. Here a coupled-channel analysis in the framework of Hamiltonian Effective Field Theory holds the promise to resolve the structure of this resonance \cite{HockleyHEFTpaper}.

\begin{acknowledgments}
	LH would like to thank Ryan Bignell for discussions and guidance in the early stages of this work. This work was supported by an Australian Government Research Training Program Scholarship and with supercomputing resources provided by the Phoenix HPC service at the University of Adelaide. This research was undertaken with the assistance of resources from the National Computational Infrastructure (NCI), provided through the National Computational Merit Allocation Scheme, and supported by the Australian Government through Grant No. LE190100021 via the University of Adelaide Partner Share. This research was supported by the Australian Research Council through Grants No. DP190102215 and DP210103706 (DBL), and DP230101791 (AWT). WK was supported by the Pawsey Supercomputing Centre through the Pawsey Centre for Extreme Scale Readiness (PaCER) program.
\end{acknowledgments}

\appendix
	
\section{Spin projection} \label{app:SpinProj} 
We use a standard interpolating field for the $ \Delta^{++} $ given by
\begin{equation}
	\chi^{\Delta^{++}}_\mu (x) = \epsilon^{abc} \qty( u^{Ta}(x) C\gamma_\mu u^b(x) ) u^c(x) \,. \label{deltainterp2}
\end{equation}
This operator has overlap with both spin-1/2 and spin-3/2 states so we need to perform spin projection to guarantee we extract the masses of the desired states. The spin-1/2 and spin-3/2 projection operators are given by \cite{Benmerrouche:1989uc, Zanotti:2003fx}

\begin{equation}
	P^{3/2}_{\mu\nu} (p) = g_{\mu\nu} - \frac{1}{3}\gamma_\mu\gamma_\nu - \frac{1}{3p^2}(\gamma\cdot p\ \gamma_\mu p_\nu + p_\mu \gamma_\nu\ \gamma\cdot p)\,, \label{proj32}
\end{equation}
\begin{equation}
	P^{1/2}_{\mu\nu} (p) = g_{\mu\nu} - P^{3/2}_{\mu\nu}(p)\,. \label{proj12}
\end{equation}

Although Eq.~(\ref{proj32}) looks somewhat cumbersome, we can do a few things to simplify these operators. First, in our lattice calculations the energy eigenstates are on shell, so we have
\begin{equation}
	p = (E,\boldsymbol{p}) = (\sqrt{\boldsymbol{p}^2 + m^2},\boldsymbol{p})\,.
\end{equation}

We consider the particles at rest, $ \boldsymbol{p} = 0 $, in which case we get $ p_0 = m $ and as usual $ p^2 = m^2 $. Further, we also have
\begin{align}
	\gamma\cdot p &= \gamma^0p^0 - \boldsymbol{\gamma}\cdot \boldsymbol{p} = \gamma^0m\,, \\
	p_\mu &= m\delta_{\mu 0} = mg_{\mu 0}\,.
\end{align}

With these results in mind, we simplify our projection operator to
\begin{align}
	P^{3/2}_{\mu\nu} (\boldsymbol{p}=0)
	&= g_{\mu\nu} - \frac{1}{3}\gamma_\mu\gamma_\nu - \frac{1}{3}(\gamma_0 \gamma_\mu g_{\nu 0} + g_{\mu 0} \gamma_\nu \gamma_0)\,. \label{reduced}
\end{align}

One can then show by using the properties of the $ \gamma $-matrices and the metric, that the elements of the projectors obey
\begin{align}
	&P^{3/2}_{0 0} (\boldsymbol{p}=0) = 
	P^{3/2}_{0 n} (\boldsymbol{p}=0) = 
	P^{3/2}_{m 0} (\boldsymbol{p}=0) = 0\,, \\
	&P^{3/2}_{m n} (\boldsymbol{p}=0) = g_{m n} - \frac{1}{3}\gamma_m\gamma_n \,, \label{res1}
\end{align}
where $ m,\, n $ are spatial Lorentz indices.

We get the corresponding results for the spin-1/2 projection operator by making use of Eq.~(\ref{proj12}) and (\ref{res1}) to obtain
\begin{align}
	&P^{1/2}_{0 0} (\boldsymbol{p}=0) = \mathbb{I}\,, \\
	&P^{1/2}_{0 n} (\boldsymbol{p}=0) =
	P^{1/2}_{m 0} (\boldsymbol{p}=0) = 0\,, \\
	&P^{1/2}_{m n} (\boldsymbol{p}=0) = \frac{1}{3}\gamma_m\gamma_n\,. \label{res2}
\end{align}

With the spin projection operators in hand, a spin-$ s $ projected correlation function is then given by
\begin{equation}
	\mathcal{G}^s_{\mu\nu} = \sum_{\sigma,\lambda = 1}^{4} \mathcal{G}_{\mu\sigma} g^{\sigma\lambda} P^s_{\lambda\nu}\,.
\end{equation}

This spin projection is performed prior to the parity projection and trace in Eq.~\eqref{trace_corr}. Thus we can obtain results for both spin-1/2 and spin-3/2 $ \Delta^+ $ states.

\section{Covariance Matrix Method} \label{app:jack}
Consider a single sample of one of the matrix elements of $ G^\pm_{ij}(t) $, which are generated by a Monte Carlo process. Call this sample $ C_{k}(t) $ where we recognise this as the $ k $-th configuration out of $ N_{con} $ total configurations. We can construct the ensemble average in the usual way
\begin{equation}
	\overline{C}(t) = \frac{1}{N_{con}} \sum_{k = 1}^{N_{con}}\, C_k(t)\,.
\end{equation}
For convenience we leave all time dependence implicit from now on and just note that the subsequent process is to be applied to each time slice.

We now introduce the jackknife estimator for the mean and uncertainty in the mean. This is a necessary step since an individual Mone Carlo sample is not an estimate of the mean. We should instead look to use averaged quantities when calculating uncertainties. 

The $ i $-th first order jackknife sub-ensemble is constructed by
\begin{equation}
	\overline{C}_i = \frac{1}{N_{con} - 1} \sum_{\substack{k = 1 \\ k\neq i}}^{N_{con}}\, C_k = \frac{N_{con}\overline{C} - C_i}{N_{con} - 1}\,.
\end{equation}
This is motivated by considering the average of $ N_{con}-1 $ samples after removing a single configuration.

We can then estimate the population mean by taking an average of the $ N_{con} $ jackknife sub-ensembles
\begin{equation}
	\overline{\overline{C}} = \frac{1}{N_{con}} \sum_{k = 1}^{N_{con}}\, \overline{C}_k\,.
\end{equation}

The standard deviation is then computed in terms of the jackknife sub-ensemble averages and the average of these,
\begin{equation}
	\sigma^2_C = \frac{N_{con} - 1}{N_{con}} \sum_{k=1}^{N_{con}} \qty(\overline{C}_k - \overline{\overline{C}})^2\,. \label{sigmac}
\end{equation}

One can reduce this to the usual expression for the standard deviation of normally distributed data by making the assumption that each $ C_k $ is a good approximate of $ \overline{C} $ and thus arrive at
\begin{equation}
	\sigma^2_C = \frac{1}{N_{con}}\frac{1}{N_{con} - 1} \sum_{k=1}^{N_{con}} \qty(C_k - \overline{C})^2\,.
\end{equation}
By taking a jackknife estimate of the samples, we are using sub-ensemble means $ \bar{C}_k $ which are $ (N-1)^2 $ times more accurate than the $ C_k $ when estimating the population mean $ \overline{C} $. 

Up until now we have disregarded any correlation between configurations on different time slices. Since the importance sampling of the lattice action forms dependencies between the time slices, configurations on nearby time slices are in fact correlated. If we wish to perform a fit to $ \overline{C}(t) $ over a time-interval, then we need a more nuanced approach to account for these cross correlations.

We hence introduce the covariance matrix which is formed by considering the deviation of the jackknife ensemble average from the average of the jackknife ensemble averages across two different time slices. Explicitly,
\begin{align}
	V(t_i,t_j) &= \frac{N_{con} - 1}{N_{con}} \nonumber \\ & \times \sum_{k = 1}^{N_{con}} \qty(\overline{C}_k(t_i) - \overline{\overline{C}}_k(t_i))  \qty(\overline{C}_k(t_j) - \overline{\overline{C}}_k(t_j))
\end{align}
For a fit window of $ T_f $ points, this provides us with a $ T_f\times T_f $ matrix, with entries along the diagonal $ V(t_i,t_i) $ equal to $ \sigma^2_c(t_i) $ using Eq.~\eqref{sigmac}. In the absence of correlations the off-diagonal factors are independent and vanish in the sum. 

We can finally compute the $ \chi^2 $ value for a fit $ T(t) $ to the mean, given by
\begin{equation}
	\chi^2 = \sum_{t_i,\, t_j} 
	\qty(\overline{C}(t_i) - T(t_i))\, 
	V^{-1}(t_i,t_j)\, 
	\qty(\overline{C}(t_j) - T(t_j))
\end{equation}
where the $ t_i $ and $ t_j $ range over all values in the interval of $ T_f $ points. The inverse of the covariance matrix is computed via the singular value decomposition algorithm. 

Finally, we obtain the degrees of freedom by counting the number of time slices in the fit interval $ T_f $, and subtracting both the number of parameters in the theoretical model and the number of singular values found when inverting the covariance matrix $ V(t_i,t_j) $.

\bibliography{bibliography}
		
\end{document}